\documentclass[preprint]{aastex}

\usepackage{emulateapj5}

\newcommand{\SIRTF}{{\sl SIRTF}}

\newcommand{\HST}{{\sl HST}}

\newcommand{\Msun}{\mbox{$M_{\sun}$}}
\newcommand{\Lsun}{\mbox{$L_{\sun}$}}
\newcommand{\Rsun}{\mbox{$R_{\sun}$}}
\newcommand{\Mjup}{\mbox{$M_{Jup}$}}

\newcommand{\perone}{\mbox{$^{-1}$}}

\newcommand{\etal}{et al.}
\newcommand{\eg}{e.g.}
\newcommand{\ie}{i.e.}

\newcommand{\kms}{\hbox{km~s$^{-1}$}}

\newcommand{\htwoo}{{\hbox{H$_2$O}}}   

\newcommand{\Ks}{\mbox{$K_S$}}
\newcommand{\degs}{\mbox{$^{\circ}$}}

\newcommand{\Teff}{\mbox{$T_{\rm eff}$}}
\newcommand{\Rc}{\mbox{${R}$}}

\newcommand{\Lp}{\mbox{${L^\prime}$}}
\newcommand{\Mp}{\mbox{${M^\prime}$}}
\newcommand{\KmLp}{\mbox{${\Ks\!-\!L^\prime}$}}
\newcommand{\JHK}{\mbox{$J\!H\!K$}}
\newcommand{\JHKs}{\mbox{$J\!H\!\Ks$}}
\newcommand{\JHKL}{\mbox{$J\!H\!K\!L$}}
\newcommand{\JHKLp}{\mbox{$J\!H\!K\!\Lp$}}
\newcommand{\JHKsLp}{\mbox{$J\!H\!\Ks\Lp$}}

\newcommand{\MKs}{\mbox{$M_{K_S}$}}
\newcommand{\Rstar}{\mbox{$R_{\star}$}}

\newcommand{\KmLpo}{\mbox{${(\Ks\!-\!L^\prime)_0}$}}
\newcommand{\EKmLp}{\mbox{${E(\Ks\!-\!L^\prime)_0}$}}
\newcommand{\cEKmLp}{\mbox{${{\cal E}(\Ks\!-\!L^\prime)_0}$}}

\slugcomment{{\em Astrophysical Journal}, in press}

\shorttitle{Disks around Young Brown Dwarfs}
\shortauthors{Liu et al.}

\begin{document}

\title{A Survey for Circumstellar Disks Around Young Substellar Objects}


\author{\sc Michael C. Liu\altaffilmark{1}}
\affil{Institute for Astronomy, University of Hawai`i, 2680 Woodlawn
Drive, Honolulu, HI 96822} 
\altaffiltext{1}{Beatrice Watson Parrent Fellow.}
\email{mliu@ifa.hawaii.edu}

\author{\sc Joan Najita} 
\affil{National Optical Astronomy Observatory, 950 North Cherry Avenue,
Tucson, AZ 85719}

\author{\sc Alan T. Tokunaga}
\affil{Institute for Astronomy, University of Hawai`i, 2680 Woodlawn
Drive, Honolulu, HI 96822}

\begin{abstract}
We have completed the first systematic survey for disks around
spectroscopically identified young brown dwarfs and very low mass stars.
For a sample of 38 very cool objects in IC~348 and Taurus, we have
obtained \Lp-band (3.8~\micron) imaging with sufficient sensitivity to
detect objects with and without disks.  The sample should be free of
selection biases for our purposes.  Our targets span spectral types from
M6 to M9.5, corresponding to masses of $\sim$15 to 100~\Mjup\ and ages
of $\lesssim$5~Myr based on current models. None appear to be binaries
at 0\farcs4 resolution (55--120~AU).
Using the objects' measured spectral types and extinctions, we find that
most of our sample ($77\%\pm15\%$) possess intrinsic IR excesses,
indicative of circum(sub)stellar disks. Because the excesses are modest,
conventional analyses using only IR colors would have missed most of the
sources with excesses.  Such analyses inevitably underestimate the disk
fraction and will be less reliable for young brown dwarfs than for
T~Tauri stars.
The observed IR excesses are correlated with H$\alpha$ emission,
consistent with a common accretion disk origin.
In the same star-forming regions, we find that disks around brown dwarfs
and T~Tauri stars are contemporaneous; assuming coevality, this
demonstrates that the inner regions of substellar disks are at least as
long-lived as stellar disks and evolve slowly for the first $\sim$3~Myr.
The disk frequency appears to be independent of mass.  However, some
objects in our sample, including the very coolest (lowest mass) ones,
lack IR excesses and may be diskless.
The observed excesses can be explained by disk reprocessing of starlight
alone; the implied accretion rates are at least an order of magnitude
below typical values for classical T~Tauri stars.  The observed
distribution of IR excesses suggests inner disk holes with radii of
$\gtrsim$2\Rstar, consistent with the idea that such holes arise from
disk-magnetosphere interactions.
Altogether, the frequency and properties of young circumstellar disks
appear to be similar from the stellar regime down to the substellar and
planetary-mass regime. This provides prima facie evidence of a common
origin for most stars and brown dwarfs.
\end{abstract}
\keywords{stars: low-mass, brown dwarfs --- infrared: stars}

\section{Introduction}

Brown dwarfs are now being found in abundance.  In the solar
neighborhood, the advent of very wide-field sky surveys have led to the
discovery of hundreds of nearby ultracool dwarfs, many of which are
likely to be substellar \citep[e.g.][]{1999ApJ...519..802K,
1999A&AS..135...41D, burg01, 2002astro.ph..4065H}.  Discoveries of more
distant substellar objects are increasing as deep, blank sky surveys
become more powerful \citep[e.g.][]{1999A&A...349L..41C,
2002ApJ...568L.107L}.  Likewise, surveys of nearby open clusters, in
particular the Pleiades, have been very successful at identifying brown
dwarfs \citep[e.g.][]{1998A&A...336..490B, 1998ApJ...504..805S,
2001A&A...367..211M}.  Searches of young ($\lesssim$10~Myr) star-forming
regions have also been quite rewarding.  Substellar objects are hotter
and more luminous in their youth, and hence it is feasible to probe to
very low masses in star-forming regions
\citep[e.g.][]{2000ApJ...540.1016L, 2000ApJ...541..977N,
2000Sci...290..103Z, 2001MNRAS.326..695L, liu01a}.  Though mass
determinations for such young objects are necessarily dependent on
theoretical models, the coolest objects seem to be $\lesssim$15~\Mjup,
comparable in mass to planets around old solar-type stars found from
radial velocity studies \citep[e.g.][]{1998ARA&A..36...57M}.

While the number of known brown dwarfs is growing rapidly, the origin of
these objects is an unanswered question.  The initial mass function
(IMF) measured in young clusters seems to be continuous from the stellar
to substellar regime \citep[e.g.][]{2000ApJ...540..236H,
2000ApJ...541..977N}, suggesting brown dwarfs form in the same manner as
stars.  However, the overlap in mass between young ``planetary-mass''
brown dwarfs ($\lesssim$15~\Mjup) and planets discovered by radial
velocity surveys is intriguing, maybe suggestive of a common genesis in
disks around young stars.  However, it is unlikely such brown dwarfs
could form like giant planets are believed to, namely core accretion
followed by runaway growth \citep{1996Icar..124...62P}.  Perhaps
gravitational disk instabilities \citep{1998ApJ...503..923B,
2000ApJ...540L..95P} or disk-disk collisions \citep{1998Sci...281.2025L,
1998MNRAS.300.1205W, 1998MNRAS.300.1214W} are involved, followed by
ejection into the field.  Indeed, a few solar-type stars are known to
have massive substellar companions \citep{els01, 2001ApJ...555..418M,
2001astro.ph.12407L, 2002ApJ...567L.133P, 2002astro.ph..2458U}, which
might point to concurrent formation of brown dwarfs and planets.  An
alternative formation scenario involving ejection has been proposed by
\citet{2001AJ....122..432R}.  They suggest that brown dwarfs form as
stellar embryos in small newborn multiple systems, but their growth is
prematurely truncated by ejection due to dynamical interactions at very
young ages ($\sim$10$^4$~yr).  A recent computer simulation of cloud
fragmentation by \citet{2002MNRAS.332L..65B} portrays an amalgam of these
scenarios: most brown dwarfs form via gravitational instabilities in
circumstellar disks and then are ejected before they can accrete to
stellar masses.

The abundance of isolated brown dwarfs in star-forming regions and open
clusters stands in stark contrast to the dearth of 20--70~\Mjup\ objects
found as close radial velocity companions to mature, solar-type stars,
a.k.a.\ the ``brown dwarf desert'' \citep{1988ApJ...331..902C,
1989ApJ...344..441M, 1995Icar..116..359W, 2000A&A...355..581H, zuck01,
2001A&A...372..935P}.  But the implication of this phenomenon is
unclear.  It might suggest that isolated brown dwarfs and massive
planetary companions arise from independent formation pathways.
Alternatively, the difference might stem from evolutionary effects which
preferentially deplete brown dwarf companions, such as inward migration
into the central star \citep{2002MNRAS.330L..11A} or outward ejection.

One potential insight into the formation mechanism(s) for brown dwarfs
is whether young substellar objects possess circumstellar disks.  There
is abundant observational evidence for accretion disks around young
solar-type stars.  Moreover, disks are a key element in current
theoretical understanding of star and planet creation: the formation of
a disk is believed to be an inevitable step, since the infalling natal
material has angular momentum.  Thus, the presence of disks around young
brown dwarfs is naturally accommodated in ``star-like'' formation
scenarios.  On the other hand, scenarios involving collision and/or
ejection are likely to be hostile to circumstellar disks.  Hence,
determining the frequency and physical properties of disks around young
brown dwarfs, as well as their relation to disks around young stars, is
an important observational goal.

Evidence for circumstellar disks around individual young ($\lesssim$few
Myr) brown dwarfs has recently been found.  Strong H$\alpha$ emission,
likely due to ongoing accretion, has been seen from sources in Taurus
\citep{1998AJ....115.2074B, 2001ApJ...561L.195M}, IC~348
\citep{1999ApJ...525..466L}, $\rho$~Oph \citep{1997ApJ...489L.165L}, and
$\sigma$~Ori \citep{2001A&A...377L...9B, 2002ApJ...569L..99Z,
2002A&A...384..937Z}.  Direct kinematic evidence for accretion from
asymmetric H$\alpha$ emission line profiles has been found for one
object in Taurus at the substellar mass limit
\citep{2000ApJ...545L.141M}.  Infrared (IR) excesses have been detected
from \JHK\ (1.1$-$2.4~\micron) data for a few spectroscopically
identified young brown dwarfs in $\rho$~Oph \citep{1999AJ....117..469W,
2000AJ....119.3019C}, IC~348 \citep{1999ApJ...525..466L}, and the
Trapezium cluster \citep{1998AJ....116.1816H, 2001MNRAS.326..695L}.  A
recent \JHK\ survey by \citet{2001ApJ...558L..51M} finds a majority of
photometrically selected brown dwarf candidates in Trapezium have
excesses.  

IR excesses are one of the classic signatures for disks around T~Tauri
stars.  Such excesses are correlated with UV/optical excess emission, as
expected for an accretion disk origin \citep{1990ApJ...354L..25H,
1993AJ....106.2024V, 1995RMxAC...1..309E, 1995ApJS..101..117K}.  In
particular, there is a one-to-one correspondence between IR color
excesses and optical veiling indicative of accretion onto the central
source \citep{1995ApJ...452..736H}.  There is other plentiful evidence
for disks around T~Tauri stars, including far-IR photometry, millimeter
wavelength studies, and high-resolution optical/IR imaging --- for
nearly all objects, the different disk indicators agree, which supports
the accuracy of using a single indicator, in this case IR excess.  On
physical grounds, disks are the best explanation for both the IR
excesses and the relatively low extinctions to the central sources
\citep[e.g.][]{1986ApJ...308..836A, 1987ApJ...319..340M}.  This body of
evidence naturally explains the IR excesses detected from young brown
dwarfs as originating from circumstellar disks.

Mid-IR (7$-$14~\micron) detections are available for young brown dwarfs
in $\rho$~Oph \citep{1998A&A...335..522C, 2001A&A...372..173B,
2002ApJ...571L.155T, 2002A&A...393..597N} and Chamaeleon
\citep{2000A&A...359..269C, 2001A&A...376L..22N, 2002ApJ...573L.115A},
which further support the disk hypothesis.  The mid-IR fluxes are
$\gtrsim$10$\times$ greater than expected from the photosphere, and the
resulting spectral energy distributions (SEDs) are much broader than can
be explained by a non-disk origin.

However, it is difficult to determine the {\em frequency} of disks
around brown dwarfs from studies to date due to a combination of small
number statistics, sample selection inhomogeneity, and, most
importantly, choice of wavelength.  A priori, brown dwarf disks are
expected to be harder to detect than disks around stars because of lower
contrast. Substellar objects are less luminous and have shallower
gravitational potentials; hence, the inner regions of their disks are
likely to be cooler and thus could have negligible excesses in the \JHK\
bands, which have been used by most previous studies.  Disks are more
luminous at longer wavelengths, and they can be readily identified at
mid-IR wavelengths. But because such observations can only detect young
brown dwarfs with large excesses, a complete census of the disk
frequency is not yet possible in the mid-IR.

\Lp-band (3.8~\micron) observations are an excellent means to address
this question.  Disks can be readily identified from ground-based
telescopes by excess \Lp-band emission, which arises from warm
circumstellar material within a few stellar radii ($\lesssim0.1$~AU).
In addition, existing infrared cameras can detect young brown dwarf
photospheres, and hence the absence of a disk can be discerned.  Such
observations have been used to measure the disk frequency for low-mass
stars in young clusters (\eg, Haisch \etal\ 2000).  These studies find a
higher disk frequency than \JHK\ or H$\alpha$ studies, as expected from
the greater sensitivity of thermal IR observations to less luminous
disks.

In this paper, we present results from a large \Lp-band survey for
circumstellar disks around young brown dwarfs and very low mass (VLM)
stars in the IC~348 and Taurus star forming regions.  The incidence and
properties of disks around young substellar objects may shed light on
the origin of these intriguing objects.  Our study targets young sources
which have been {\em spectroscopically classified} to be very cool, with
spectral types of M6 and later.  By focusing on targets with spectral
types, we are more sensitive to small IR excesses since we can determine
the intrinsic photospheric colors.  Our selection criterion also ensures
that we are targeting very low mass members, lying near or below the
stellar/substellar mass boundary.  Since our observations are sensitive
enough to detect bare photospheres, this is the first unbiased survey
for disks around such objects.

We describe our sample and our observations in \S~2. We discuss the
frequency of IR excesses and correlations with other properties in \S~3.
Simple models are used to study the disk characteristics in \S~4.  We
examine the possible systematic effects in \S~5.  We then discuss the
implications of our findings for brown dwarf disk searches, the
properties of such disks, and brown dwarf formation scenarios in \S~6.
Finally, we summarize our results in \S~7.  For readers primarily
interested in the results, we suggest focusing on \S~6 and \S~7.


\section{Observations}

\subsection{Sample Selection and Properties \label{sec:sample}}

For objects in young star-forming regions, the substellar boundary is
believed be around spectral type M6, based on comparison with
theoretical models \citep{1998ApJ...493..909L} and the application of
the lithium test in young clusters \citep{1998bdep.conf..394B}.  Most
young brown dwarf candidates do not have lithium measurements, which
would guarantee their status as cluster members.  However, their
positions in color-magnitude diagrams indicate they are very likely
cluster members and the very late spectral types are consistent with
substellar status.

Our sample of 38 objects includes nearly all published objects with
spectral types of M6 or later in the IC~348 and Taurus star-forming
regions. We adopt a distance of 140~pc to Taurus
\citep{1978ApJ...224..857E, 1994AJ....108.1872K} and 300~pc to IC~348
\citep[][and references therein]{1998ApJ...497..736H}, with an error of
10\% for each.  The typical stellar ages in these regions have been
estimated to be around 1$-$3~Myr \citep{1995ApJS..101..117K,
1998ApJ...497..736H, 2000ApJ...540..255P}.  There are three components
to our sample:
\begin{itemize}

\item {\em IC 348 core region}: these objects originate from the NICMOS
imaging survey of \citet{2000ApJ...541..977N}.  By using narrow-band
photometry to measure the depth of the 1.9~\micron\ \htwoo\ absorption
band, they derived spectral types and extinctions for all objects down
to $K=16.5$~mag in the central $\approx$~5\arcmin$\times$5\arcmin\
region.\footnote{The spectral types in the IC~348 core sample are
estimated from the correlation between $I$-band spectral types and the
strength of the 1.9~\micron\ water band, as measured for both field
dwarfs and a subsample of young stars in IC~348 itself (see Najita
\etal\ 2000 for details).  The use of fractional M~subtypes reflects the
continuous nature of the water index measurement and is not meant to
advocate a new spectral typing convention.}  This sample of substellar
objects is unique because it is {\em complete}, in the sense that every
object in the survey region has an estimated spectral type.  Among their
identified cluster members, we observed nearly all objects with spectral
types from M5.7 to M9.4, with the exception of two (objects 064-05 and
014-03, with spectral type of M5.8 and M6.3, respectively).  Cluster
members were selected as objects being 5~Myr or younger in their
color-magnitude diagram. We also observed one object (071-01, spectral
type M5.8) with an estimated age of $\approx$10~Myr.  A number of these
sources have published optical spectroscopy (see~\S~\ref{sec:sptypes}),
and we have also been pursuing ground-based optical and near-IR
spectroscopy (M. Liu \etal, in preparation); the spectroscopy confirms
that these sources are genuine very cool (late M-type) young objects.
Since this NICMOS sample has been selected and analyzed in a homogeneous
fashion, our analysis will often examine the properties of this
sub-sample in addition to those of our sample as a whole.

\item {\em IC 348 outer region}: we include additional very cool members
of IC~348 from \citet{1999ApJ...525..466L}, originally selected from
\Rc\ and $I$-band optical imaging. These sources lie outside the
\citet{2000ApJ...541..977N} survey region.  \citet{1999ApJ...525..466L}
used far-red optical spectroscopy to derive spectral types and $J$-band
extinctions.  No errors are given for his extinction determinations; we
conservatively assume these are 0.1~mag.

\item {\em Taurus}: we include nearly all published objects with
spectral types of M6 or later in the Taurus star-forming region. These
objects were originally found from optical or X-ray surveys.  Four of
them come from \citet{2001ApJ...561L.195M}, who derived spectral types
from far-red optical spectroscopy and $V$-band extinctions from
broad-band colors.  The remaining objects have spectral types from
\citet{1998AJ....115.2074B} and \citet{1998ApJ...493..909L}; for these,
we use $H$-band extinctions derived by \citet{2000ApJ...544.1044L} from
broad-band colors.  No errors are given for the published extinction
determinations; we conservatively assume these are 0.5~mag and 0.1~mag
for the \citet{2001ApJ...561L.195M} and \citet{2000ApJ...544.1044L}
results, respectively.

\end{itemize}

\begin{figure}
\vskip -0.25in
\hskip -0.25in
\centerline{\includegraphics[width=5.2in,angle=0]{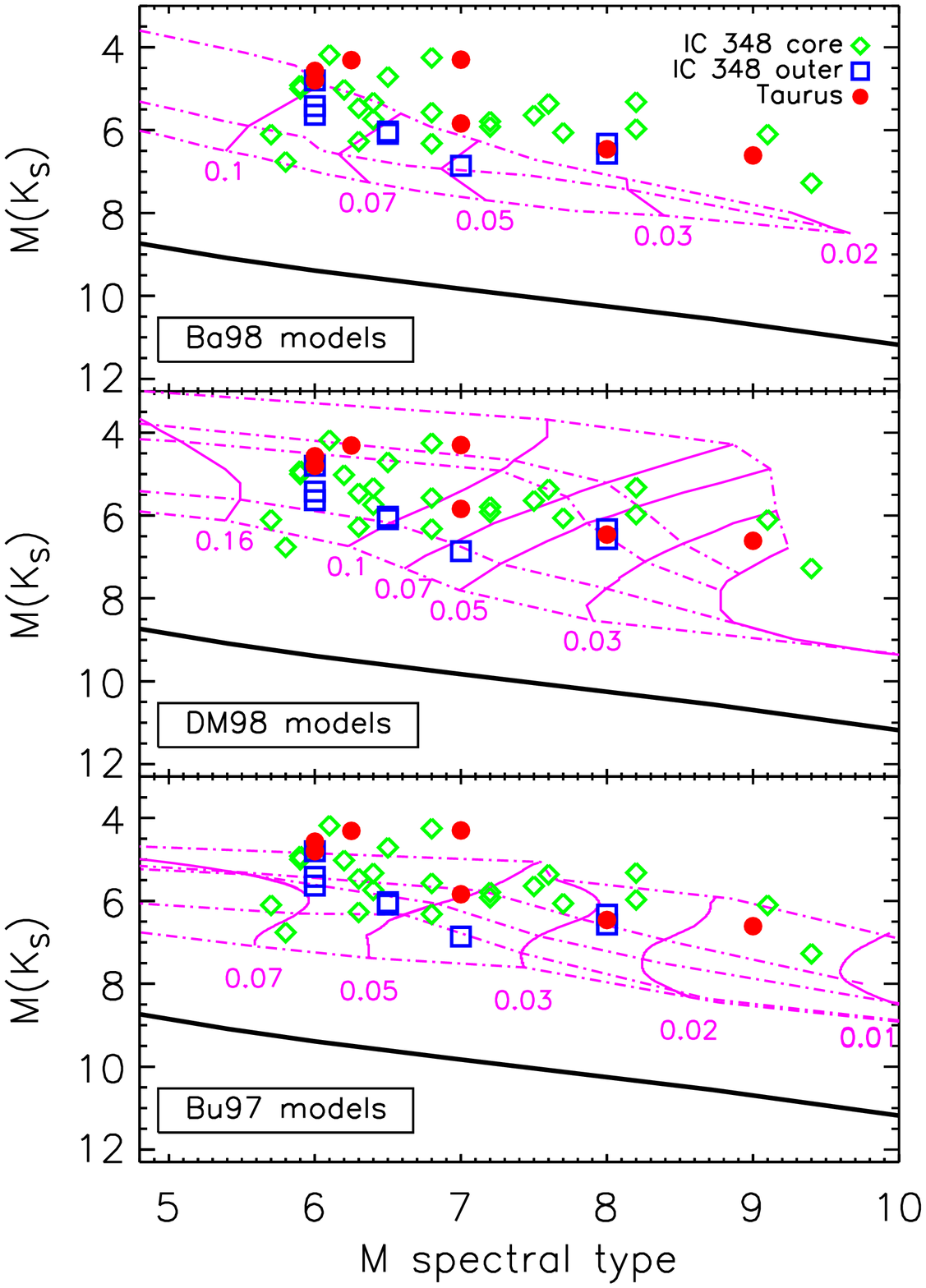}}
\vskip 6ex
\caption{\normalsize Dereddened absolute \Ks-band magnitude and spectral
type for our \Lp-band sample.  Magnitude errors are comparable to the
symbol size, and spectral type errors are $\lesssim$0.5~subclasses.  The
main-sequence is shown as a heavy line. Dotted lines are model
isochrones while solid lines show models of constant mass in units of
\Msun.  The \citet{1999ApJ...525..466L} temperature scale is used to
convert the model temperatures to spectral type.  {\bf Top:}
\citet{1998A&A...337..403B, 2002A&A...382..563B} models.  The isochrones
are for ages of 1, 5, and 10~Myr.  {\bf Middle:}
\citet{1997MmSAI..68..807D} models.  The isochrones are for ages of 0.1,
0.5, 1.0, 5.0, and 10.0~Myr. {\bf Bottom:} \citet{1997ApJ...491..856B}
models.  The isochrones are for ages of 0.1, 0.5, 1.0, 5.0, and
10.0~Myr. \label{plot-hrd}}
\end{figure}

Figure~\ref{plot-hrd} plots our targets in comparison to theoretical
models from \citet{1998A&A...337..403B, 2002A&A...382..563B},
\citet{1997MmSAI..68..807D}, and \citet{1997ApJ...491..856B}. For the
D'Antona \& Mazzitelli models, we use a newer unpublished version, which
we refer to as the DM98 models.\footnote{The models are available at
{\tt http://www.mporzio.astro.it/\~{}dantona}.}  To compare the models
to the observations, we use a linear fit for the conversion between
spectral type and effective temperature (\Teff) advocated by
\citet{1999ApJ...525..466L} for pre-main sequence objects:
\begin{equation}
\Teff (K) = 3850 - 141.0 \times SpT \label{eqn:luhmanscale}
\end{equation}
where $SpT$ is the M~spectral subclass, \ie, $SpT=0$ means M0, $SpT=5$
means M5, etc.  To plot the DM98 and Burrows \etal\ models, we also use
a linear fit to $K$-band bolometric corrections derived by
\citet{2000ApJ...535..965L, 2001ApJ...548..908L}:
\begin{equation}
BC_K = 3.99 - \Teff/2424\ .
\label{eqn:bolcorr}
\end{equation}
The scatter about the fit is $\approx$0.1~mag in $BC_K$, negligible for
our purposes.

The model-derived masses range from $\approx$15 to 100~\Mjup.  These
depend on the choice of models and temperature scale; our choices lead
to conservative (\ie, larger) values for the masses (see
\S~\ref{sec:tempscale}).  The ages inferred from the models are
$\lesssim$5~Myr, and do not depend strongly on the choice of temperature
scale since the isochrones are roughly horizontal in the HR diagram.
Taken at face value, some models indicate that the lower mass objects
tend to be younger than the higher mass ones.  However, given the
substantial uncertainties in the models of young substellar objects
\citep{2002A&A...382..563B}, we largely avoid using model-derived
quantities in our analysis.  Instead, we assume that the stars and brown
dwarfs in these regions are approximately coeval, and adopt the age
estimate of a few Myr derived for the stars.

\subsection{\Lp-band Imaging: UKIRT}

We obtained \Lp-band (3.4--4.1~\micron) imaging on 13$-$15 October 2001
UT at the 3.8-m United Kingdom Infrared Telescope (UKIRT).  We used the
facility camera IRCAM/TUFTI, which employs a $256\times256$ InSb array
and has a pixel scale of 0\farcs081~pixel\perone.  The filter used was
the \Lp-band filter produced by the Mauna Kea Filter Consortium
(3.4$-$4.1~\micron; \citealp{mkofilters1, mkofilters2}).  UKIRT has a
fast-steering secondary mirror which provides tip-tilt correction by
sensing a star off-axis of the science target. The typical resulting
image quality was excellent (0\farcs35$-$0\farcs4 FWHM), and none of our
sources were resolved into binaries.  Conditions were photometric.
During the course of the night, we regularly observed standard stars
from a UKIRT in-house calibrator list.

We used a standard ``two-beam'' observing technique in order to
accurately cancel the very large and variable thermal flux from the
atmosphere. For each target, we obtained a coadded set of short
integrations, then offset the telescope by about half of the array and
integrated again.  This was repeated, with small dithers at each nod
position to ensure the object did not fall on the same position on the
array. Integration times were determined using the published $K$-band
magnitudes, spectral types, and extinctions to reach comparable S/N for
all the objects.  The on-source integration ranged from 3 to 60~min.

Reductions were done in standard fashion for thermal IR imaging.  For
each object, a flat-field was created from the average of all the
images.  Each flat-fielded image was then subtracted from the
corresponding image in the other nod position.  The images were
registered using the relative telescope coordinates and averaged
together.  The resulting mosaic contained a positive and negative image
of the target. Both were measured, and the standard error of the results
was adopted as the uncertainty.  For the standard stars, we used a
5\arcsec\ radius, as is done for the UKIRT standard star programme.  We
then determined a zeropoint and extinction coefficient for each night.
Since most of our targets are faint, we used a smaller aperture (10
pixel radius) for their photometry, and then applied an aperture
correction out to 5\arcsec\ based on curves of growth measured nightly
from the much brighter standard stars. The size of this correction was
on the order of 0.15~mags.

Table~\ref{mags-final} presents our resulting measurements.  The final
errors for the object magnitudes are the quadrature sum of the errors in
the raw photometry (the dominant term), the zeropoint, the extinction
coefficient, the aperture correction, and the standard star magnitudes.

One object in the photometry sample, 045-02 in IC~348, was significantly
bluer in \KmLp\ than the colors of late M~dwarfs.  Based on \JHKLp\
colors, it is probably a background late-type giant whose position in
the observed color-magnitude diagram overlaps the cluster members.  We
include its photometry in the Table~\ref{mags-final}, but the object is
excluded from the analysis.

\subsection{IR Photometry: 2MASS \label{sec:2mass}}

We applied very small transformations to all the relevant photometry,
namely the IR colors for field and young cluster M~dwarfs, to bring them
to a common photometric system.  We chose to use the well-understood
2MASS system \citep{2001AJ....121.2851C}.  For the \JHK\ colors of
M~dwarfs in the solar neighborhood, we used measurements from
\citet[][hereafter L92]{1992ApJS...82..351L}, \citet[][hereafter
L98]{1998ApJ...509..836L}, and \citet[][hereafter L02]{leg01}, excluding
the subdwarfs.  The L92 and L98 near-IR data are on the CIT system,
which we converted to the 2MASS system using
\citet{2001AJ....121.2851C}.  The data from L02 are on the MKO-NIR
system, which we converted using additional transformations from
\citet{2001MNRAS.325..563H}.  The transformed samples showed no
significant differences in their near-IR colors.  For the \Lp-band data,
all the data were either in the UKIRT system (L92 and L98 data) or the
MKO-NIR system (L02 data and our observations).  These are essentially
identical for M~dwarfs so no transformations were needed (L02).

The \citet{2000ApJ...541..977N} sample has been observed in the
HST/NICMOS F166N and F215N filters, which are 1\% bandwidth.  These
filters are placed in feature-free regions of $H$- and $K$-band stellar
spectra.  To convert these measurements to a standard photometric
system, we matched 77 objects in the Najita \etal\ sample with objects
in the 2MASS Second Incremental Data Release \citep{2mass-2}.  We found
(F166N--$H$) = 0.00$\pm$0.02~mags and (F215N--$\Ks$) =
0.08$\pm$0.01~mags with no statistically significant dependence on $J-K$
color.  We added $J$-band data from 2MASS, available for 23 out of 25
objects.

For the Luhman (1999) IC~348 sample, we used 2MASS near-IR photometry,
available for all the objects except one (\#405).  For that one, we used
his published photometry.  The published near-IR photometry for the
Taurus sample all came from 2MASS.

\section{Observational Results}

We identify IR excesses around our targets using \KmLp\ colors, which
track the shape of the SEDs from 2.0--4.1~\micron.  This choice is
advantageous for several reasons, including: (1) disks are more luminous
at longer wavelengths, as compared to just using \JHK\ data; (2) the
effects of extinction (and uncertainties therefrom) are reduced at
longer wavelengths; (3) the small wavelength range minimizes the effect
of extinction errors; (4) previous studies of disks around T~Tauri stars
have used these wavelengths; (5) color measurements are distance
independent; and (6) there is ample published \KmLp\ data for field
M~dwarfs.  Note that all the results discussed in this section are {\em
strictly empirical}, \eg, independent of theoretical models and choice
of temperature scale.

\subsection{IR Excess Frequency and Amplitude}

The top panel of Figure~\ref{plot-kl} shows our observations: \KmLp\
colors for the sample as a function of spectral type compared with the
colors of the field M~dwarf sample described in \S~\ref{sec:2mass}.  A
substantial number of objects show \KmLp\ colors redder than expected
from purely photospheric emission, while the lower envelope of the
observed color distribution agrees well with the dwarf locus.

Since all objects in our sample have measured spectral types and
extinctions, we can determine their intrinsic photospheric \KmLp\
colors, and hence we are very sensitive to the presence of excess IR
emission, \ie, the presence of circumstellar disks.  The bottom panel of
Figure~\ref{plot-kl} shows the dereddened \KmLp\ colors, using the
published extinctions and the extinction law from \citet{mat00}.  A few
trends are apparent: (1) most objects have dereddened \KmLp\ colors in
excess of that expected from their photospheres alone; (2) the lower
envelope of the color distribution is consistent with the locus of field
M~dwarfs, suggesting that such objects provide a legitimate comparison
(see also \S~\ref{sec:gravity}); and (3) the maximum amplitudes of the
IR excesses decrease at later spectral types, \ie, lower masses.

For each object, we compute the intrinsic \KmLp\ excess in the usual
fashion:
\begin{equation}
\EKmLp = (\KmLp)_0^{observed} - (\KmLp)_0^{photospheric}\ .
\label{eqn:excess}
\end{equation}
The photospheric colors as a function of spectral type are determined
from the field M~dwarf sample.  To compute the error in \EKmLp, we add
in quadrature the errors in the \KmLp\ colors, the reddening, and the
spectral types (which add an uncertainty to the intrinsic photospheric
color).  The resulting excesses and their errors are listed in
Table~\ref{mags-final}.  In general, the excesses are much smaller than
those observed for T~Tauri stars \citep[e.g.][]{1995ApJS..101..117K}.

Table~\ref{table-excess} gives the breakdown of the IR excess fraction
by spectral type and target subsample.  Overall, 31 out of 38 (82\%) of
the sample show \KmLpo\ excesses, with 21 out of 38 (55\%) having
excesses larger than their 1$\sigma$ measurement error.  To compute the
disk frequency and its uncertainties, we use a Monte Carlo technique
described in Appendix~A which accounts for both the (gaussian)
measurement errors and the (Poisson) counting errors.  We find an excess
frequency of $77\%\pm15\%$ for the sample --- disks around young brown
dwarfs and VLM stars appear to be very common.

Next, we examine any correlations between the IR excesses and other
physical properties of the objects.  Note that since the detected amount
of disk emission depends on the viewing angle to the observer, any such
trends are not expected to be very strong.  Random variations in the
viewing orientation will inevitably act to obscure such trends.


\subsection{Correlation with H$\alpha$ \label{sec:halpha}}

For T~Tauri stars, both the optical line emission and the IR excess are
believed to originate from an accretion disk.  The IR excess comes from
warm dust grains in the disk, while H$\alpha$ emission originates from
the accretion of disk material onto the central star, \eg, via a
boundary layer or a magnetically regulated accretion flow.  Therefore,
the H$\alpha$ emission and IR excesses should be correlated, and indeed
such a correlation is seen among T~Tauri stars
\citep[e.g.][]{1993AJ....106.2024V, 1995ApJS..101..117K}.
Figure~\ref{halpha} shows such a comparison for our sample of young VLM
stars and brown dwarfs, using H$\alpha$ data mostly from the literature
(\citealp{1998AJ....115.2074B, 1998ApJ...493..909L, 1998ApJ...508..347L,
1999ApJ...525..466L, 2001ApJ...561L.195M}; K.~Luhman 2002, priv.\
comm.).  The entire Taurus sample has published H$\alpha$ measurements,
and about half of the IC~348 sample does.  A correlation is seen between
\EKmLp\ and H$\alpha$ equivalent width, with a Spearman rank correlation
coefficient $r_S=0.64$. The probability of this being drawn from a
random sample is 0.0009, or a 3$\sigma$ correlation.  This is comparable
to the correlation observed for T~Tauri stars
\citep{1995ApJS..101..117K} and provides strong circumstantial evidence
for accretion disks around young brown dwarfs.


\subsection{Non-Correlations with Mass and Age}

Figure~\ref{klexcess-hist} presents a histogram of \KmLpo\ excesses as a
function of spectral type.  For the three earliest spectral type bins
(M5.7$-$M6.4, M6.5$-$M7.4, and M7.5$-$M8.4), a Kolmogorov-Smirnov (K-S)
test \citep{1992nrca.book.....P} finds a high probability
($\gtrsim$40\%) that all the samples are drawn from the same parent
population.  Since the evolutionary tracks at constant mass are roughly
constant in \Teff\ for young ages, the spectral types correspond to a
relative mass scale.  Hence, we find no evidence for a strong dependence
of IR excesses on mass.  The exceptions are the coolest (lowest mass)
objects, types M8.5$-$M9.4, where the excesses are small, consistent
with non-existent.
The disk frequency likewise appears to be independent of mass
(Table~\ref{table-excess}), though larger samples of the coolest objects
are needed to better study this issue. Using only the homogeneous IC~348
core sample gives comparable results.

In Figure~\ref{klexcess-hist}, there is a hint that the Taurus
population might tend to have larger \KmLpo\ excesses than the IC~348
samples.  However, a K-S test of the M6$-$M7 objects shows the Taurus
\KmLpo\ excess distribution differs from that of the IC~348 core sample
at only the 1.5$\sigma$ level.  Also it is not clear if/how the
selection effects of the Taurus sample impact such a comparison.  Hence,
we find no strong evidence for differences in the sub-samples' IR
excesses.

Figure~\ref{agetrend} examines the age dependence of the inner disk
properties by plotting the \KmLpo\ excesses as a function of
dereddened absolute \Ks-band magnitude, \MKs.  If we separate objects by
spectral subclass, \MKs\ can be used as a surrogate for age, due to the
fact that model isochrones are roughly horizontal in the HR diagram.  We
find no statistically significant ($\gtrsim1\sigma$) correlation between
the IR excess and \MKs\ using the Spearman rank correlation test. This
lack of correlation is also seen for the more homogeneous IC~348 core
sample. (The exception is the M6.5$-$M7.5 bin, which exhibits a
correlation of modest statistical significance.  But this is driven by
the one Taurus object with a very large \EKmLp. When this object is
excluded, the correlation disappears.)
If we adopt an age for the substellar population based on that estimated
for the stars (see \S~\ref{sec:sample}), our findings suggests that the
inner regions ($\lesssim$0.1~AU) of young brown dwarf disks do not
evolve substantially over the first $\sim$3~Myr.  This timescale is in
accord with studies of disks around T~Tauri stars
\citep[e.g.][]{1989AJ.....97.1451S, 1990AJ.....99.1187S}.


\section{Analysis of Disk Properties}

\subsection{Disk Models \label{sec:models}}

We use simple models of circumstellar disks to examine the observed
infrared excesses.  Our goal here is not to deduce detailed physical
properties, but to understand the general characteristics and trends of
our observations.  For simplicity, we assume that the disks are
optically thick and geometrically flat.  Similar models have been widely
used to understand the emission from disks around the higher mass
T~Tauri stars \citep[e.g.][and references therein]{1974MNRAS.168..603L,
1987ApJ...312..788A, 1992ApJ...393..278L, 1998apsf.book.....H}.  The
assumption of a flat disk is appropriate for the inner disk region from
which the \KmLp\ emission originates.  Although in reality disks
probably flare significantly at larger radii, the impact of the flaring
on the SED is much more important at mid-IR ($\sim$10~\micron) and
longer wavelengths \citep[e.g.][]{1987ApJ...323..714K,
2002ApJ...567.1183W}.  Indeed there are some indications from mid-IR
measurements that flat disks might be more appropriate for some young
brown dwarfs \citep{2002A&A...393..597N}.

For both active accretion and passive reprocessing of stellar
irradiation, the disk temperature profile follows $T \propto r^{-3/4}$,
where $r$ is the radial position in the disk.  The normalization of the
temperature profile is determined from the total luminosity that is
produced by reprocessing and accretion.  (The heating of the star by
radiation from disk is ignored.)  For a disk that extends from an
infinite outer radius to an inner radius $R_{in}$, the luminosity from
reprocessing is
\begin{equation}
L_{rep} = 0.25 {\left(R_* \over R_{in}\right)} L_*
\label{eqn:lumrep}
\end{equation}
where $L_*$ and $R_*$ are the luminosity and radius of the central
source, respectively \citep{1986ApJ...308..836A}.  For the same disk,
the luminosity from accretion is given by
\begin{equation}
L_{acc} = {GM_*\dot M\over 2R_{in}}\ ,
\label{eqn:lumacc}
\end{equation}
where $M_*$ is the mass of the central source and $\dot M$ is the disk 
accretion rate.  Half of the total potential energy of the accreting
material is radiated away, with the other half stored as rotational energy.
In principle, the accretion luminosity can become arbitrarily large,
since it scales with $\dot M$, in contrast to the reprocessing
luminosity which has a maximum of $0.25L_*$.  In practice, for typical
T~Tauri stars, the luminosity in excess of the stellar photospheres is
$\approx0.25L_*$, \ie, about the maximum expected from a flat
reprocessing disk alone \citep{1998apsf.book.....H}.

Given the contributions from reprocessing and accretion, we compute the
resulting disk SED by summing over annuli, assuming each annulus
radiates as a blackbody.  To compute broad-band magnitudes from the
resulting model SEDs, we use transmission profiles which account for the
combined spectral response of the detector, filter, and atmosphere.  We
obtained $J\!H\!\Ks$ profiles from 2MASS, and \Lp\Mp\ ones for the MKO
system from UKIRT (S. Leggett, priv. comm.).  In computing the emergent
flux from the star+disk spectrum, we account for the non-blackbody
nature of the stellar spectra, a small, but noticeable, effect for cool
M-type photospheres.  We adjust the stellar spectra in the models to
match the IR color locus for field M~dwarfs described in
\S~\ref{sec:2mass}.  In doing so, we make use of bolometric corrections
computed by \citet{2000ApJ...535..965L, 2001ApJ...548..908L}.  Although
we have ignored, for simplicity, the true non-blackbody nature of the
disk emission \citep[e.g.][]{1992RMxAA..24...27C}, we note that the
results of these simple models agree well with those of Calvet \etal\ as
presented in \citet{1997AJ....114..288M} for the low accretion rates
considered here.

Table~\ref{table-passive} tabulates the maximum possible excess from our
reprocessing disks, \ie, for a face-on orientation with no inner hole.
Figure~\ref{plot-passive-2panel} shows the resulting $JH\Ks\Lp$
color-color diagram for stars with reprocessing disks for a range of
central stars, viewing angles, and hole sizes.  The resulting models all
have roughly the same color-color slope.  The plotted stellar loci for
giant and dwarf stars are from \citet{1988PASP..100.1134B}, converted to
the 2MASS system.  We use 2nd order polynomial fits to represent the
mean colors of M1 to M9.5 dwarfs, based on \citet{1992ApJS...82..351L},
\citet{1998ApJ...509..836L}, and \citet{leg01} data; the rms scatter
about the fit is consistent with the measurement errors.  We adopt the
reddening law from \citet{mat00} for $R_V=3.1$.  Notice that (1)~the
disks around earlier type (hotter, more luminous) objects generate much
larger IR excesses than those around later type (cooler, less luminous)
objects; (2)~mid-M type objects (near the stellar/substellar boundary)
with disks are predicted to be {\em bluer} in $J-H$ than either early-M
types (representative of T~Tauri stars) or late-M (substellar) objects
with disks; and (3) models with significant holes generate color
excesses in only \KmLp, because the inner disk regions are relatively
cool.


\subsection{Evidence for Inner Disk Holes}\label{sec:holes}

\Lp-band observations are sensitive to emission from the hot, inner
portions of circumstellar disks.  For the range of \Teff\ considered
here, the emission comes from a region within a few radii of the central
source ($\lesssim$0.1~AU).  Hence, the presence of inner holes in the
disks will have a substantial impact on the \KmLp\ emission.  Disks with
inner holes of a few stellar radii have been inferred to exist around
the higher mass T~Tauri stars \citep[e.g.][]{1996ApJ...462..439K,
1997AJ....114..288M} and are believed to originate from the interaction
of the stellar magnetic field with the accretion disk
\citep[e.g.][]{1988ApJ...330..350B, 1991ApJ...370L..39K,
1994ApJ...429..781S}.  Modeling and interpreting the disk emission from
a specific object depends on the viewing angle to the observer, which is
unknown.  This limitation can be overcome with a large unbiased sample
of objects, as is the case for our study --- meaningful constraints on
the inner holes can then be found from (1)~the maximum observed IR
excess and (2) the observed distribution of IR excesses.

Figure~\ref{plot-kl-holes} illustrates the effect of inner holes on
passive disk models viewed face-on.  Holes of $\gtrsim$4\Rstar\ are
sufficient to quench any \KmLp\ excess, especially for later-type
(cooler) objects.  One interesting aspect of Figure~\ref{plot-kl-holes}
is that none of our targets are observed to have an IR excess as large
as the model without an inner hole ($R_{in} = R_*$).  The observed upper
envelope is more consistent with the $R_{in} \approx 2\!-\!3\Rstar$
models, suggesting inner holes are common.  Note that for a specified
hole size, {\em the distribution of excesses should peak near the
maximum excess} based on simple geometrical considerations.  The
probability distribution of viewing angles follows $P(\theta) = \sin
\theta$.  Although disks seen nearly face-on ($\theta\sim0\degs$) are
rare, the decrease of disk flux with $\cos\theta$ compensates for this.
The net result is that the color distribution in magnitudes for a random
population of disks is strongly peaked towards the maximum excesses
\citep[e.g.][]{1990ApJ...349..197K}.

To quantitatively examine the possibility of inner holes, we compare the
observed \EKmLp\ color distribution against Monte Carlo realizations of
star+disk models viewed at randomly selected viewing angles.  We first
construct a series of passive disk models with a fixed inner radius.
The amplitude of the IR excess depends on the viewing angle $\theta$ and
the \Teff\ of the central source: hotter stars with disks can produce
larger IR excesses than cooler stars with disks. However for the range of
temperatures considered here, the \Teff\ dependence is removed when the
model excesses are normalized against the maximum possible excess (\ie,
$\theta=0\degs$).  This normalization allows us to examine models which
depend only on the viewing angle, which are needed to compare with the
Monte Carlo simulations.  Thus, for each object, we compute the {\em
normalized excess}, defined as
\begin{equation}
{\cal E}(\KmLp)_0 = {E_{K_S-L^\prime}^{obs} 
   \over \max(E_{K_S-L^\prime}^{model})}
\label{eqn:normexcess}
\end{equation}
where the numerator is the observed \KmLpo\ excess and the denominator is
the \KmLp\ excess from a face-on passive disk model computed for the
\Teff\ corresponding to the object's spectral type.  \cEKmLp\ is simply
the ratio of the observed excess to the maximum possible model excess.

We exclude objects with small observed \EKmLp\ from this analysis; these
objects would correspond to model disks viewed nearly edge-on.  In such
cases, the disk probably obscures the central source, and hence these
objects are missing from the original imaging surveys.  Objects in our
sample with small \EKmLp\ are probably not those with disks viewed
nearly edge-on, but rather objects without significant disks.  We choose
a cutoff value of $\theta=80\degs$, corresponding to a 100~AU flared
disk with an outer height of 17~AU; this agrees well with results from
\HST\ imaging of the edge-on Taurus object HH~30
\citep{1996ApJ...473..437B}.

For a given inner hole radius, we compare the observed \cEKmLp\
distribution with that expected from a Monte Carlo set of disks chosen
with random viewing angles.  The results are shown in Figure~\ref{holes}
for models with inner disk radii of 1, 2, and 3\Rstar.  Models without
inner holes predict many more objects with large values of \cEKmLp\ than
observed; this is equivalent to the findings in
Figure~\ref{plot-kl-holes}, where there are no objects with excesses as
large as the face-on models without holes.  On the other hand, models
with inner radii of 3\Rstar\ predict much smaller excesses than
observed.  Only models with inner radii of $\approx2\Rstar$ agree well
with the data.  This is confirmed using the K-S test: for a hole radius
of 2.2\Rstar, the K-S test indicates a 82\% chance that the observed and
the predicted \cEKmLp\ distributions originate from the same population.
This is much better concordance than for the 1\Rstar\ and 3\Rstar\ hole
models; the K-S test indicates a $\gtrsim1000\times$ lower probability
of these models being drawn from the same parent population as the
observations.

To summarize, we find the observed maximum amplitude and color
distribution of the \KmLpo\ excesses point to disks with characteristic
inner radii of $\approx2\Rstar$.  Note that our analysis based on
passive disks provides a {\em lower estimate} of the hole size: larger
holes could be accommodated if the luminosity from accretion is
substantial (see \S~\ref{sec:accretion}) or if the vertical flaring of
the disks is unexpectedly large.  Mid-IR observations of young brown
dwarfs in $\rho$~Oph may also suggest the presence of inner holes
\citep{2002A&A...393..597N}.  Finally, as illustrated in
Figure~\ref{plot-kl-holes}, holes of $\gtrsim2\Rstar$ could account for
the small/negligible IR excesses observed for the M9 sources in our
sample.  Longer wavelength data are need to conclusively determine
whether these very cool objects have disks.


\subsection{Constraints on Accretion Rates}\label{sec:accretion}

The widespread presence of H$\alpha$ emission among the sample
(\S~\ref{sec:halpha}) indicates that accretion is ongoing for most
objects.  To roughly constrain the accretion rate, we rearrange
equations~(\ref{eqn:lumrep}) and~(\ref{eqn:lumacc}) to compare the
relative contributions of reprocessing and accretion luminosity for the
case of a simple flat accretion disk:
\begin{eqnarray}
{L_{acc} \over L_{rep}} & = & 
   0.3 \left(\dot M \over {10^{-9}\, \Msun/{\rm yr}}\right)
       \left(M_{BD} \over {60\, \Mjup}\right) \times \nonumber \\
  & &  \qquad \left(L_{BD} \over {0.02\, \Lsun}\right)^{-1}
       \left(R_{BD} \over {0.6\, \Rsun}\right)^{-1}
\label{eqn:lumacclumrep}
\end{eqnarray}
where we have used values for the mass, luminosity and radius
representative of our sample.  As discussed in the previous section, the
magnitude of the observed IR excesses can be generated by a reprocessing
disk alone.  There is no need to invoke significant luminosity from mass
accretion, which suggests accretion rates are
$\lesssim10^{-9}$~\Msun~yr\perone.  Such rates are comparable to the
lowest accretion rates measured for classical T~Tauri stars (CTTS);
typical CTTS rates are around 10$^{-7}$ to 10$^{-8}$~\Msun~yr\perone\
\citep[e.g.][]{1993AJ....106.2024V, 1998ApJ...492..323G}.  Such low
rates also indicate that young brown dwarfs accumulate at most
$\sim$10\% of their final mass via disk accretion, if the current rates
are representative of their time-averaged accretion history.

For one of our targets, the M6 object V410~Anon~13 in Taurus,
\citet{2000ApJ...545L.141M} find its H$\alpha$ line profile is
well-fitted using magnetospheric accretion models in which the
magnetosphere extends to 2.2$-$3\Rstar, in accord with our analysis of
the \KmLpo\ excesses in the previous section.  They derive an accretion
rate of $\sim\!5\times10^{-12}$~\Msun~yr\perone, well below our estimated
upper limit.  More such direct measurements for young brown dwarfs are
sorely needed to determine the typical accretion rate and its
dispersion.

As mentioned before, there is a degeneracy in determining the inner hole
size and the accretion rate based on the IR data alone.  Pure
reprocessing disks with $\approx$2\Rstar\ inner holes can account for
the observed IR excesses (\S~\ref{sec:holes}). But disks with larger
inner holes can also agree with the data if the accretion rates are
significant.  For instance, disk models with 3\Rstar\ holes and
accretion rates of $\sim$10$^{-9}$~\Msun~yr\perone\ can generate
comparable excesses.  However, such accretion rates would disagree with
H$\alpha$ measurements of young brown dwarfs by
\citet{2000ApJ...545L.141M, muz02}, which include a few objects common
to our \Lp-band sample.  Even larger inner holes of 4$-$5\Rstar\ would
make reprocessing nearly irrelevant (\eg, Figure~\ref{plot-kl-holes}),
and the accretion rates required to generate the observed excesses
($\gtrsim$10$^{-8}$~\Msun~yr\perone) would be even more discrepant.
Thus it is unlikely that the characteristic sizes of the inner holes are
much larger than $\approx$2\Rstar.


\section{Potential Systematic Effects \label{sec:errors}}

Our identification and analysis of IR excesses depend on several
measurements which may be prone to systematic errors.  Here, we find
that they should not have a significant impact.

\subsection{Extinctions and Spectral Types \label{sec:sptypes}}

Measuring the IR excesses depends on the adopted spectral types and
extinctions.  As a check on these, we examine the subset of our IC~348
core sample which has been studied by more than one group.  Twelve of
the objects, about half the sample, have been classified using both IR
narrow-band spectrophotometry \citep{2000ApJ...541..977N} and
optical/far-red spectroscopy \citep{1999ApJ...525..466L}; these twelve
are roughly distributed across our sample's full range of spectral
types.  The $K$-band extinctions measured by the two studies agree well,
with a median difference of 0.03~mag and an rms of 0.12~mag.  This
concurs with the analysis of Najita~\etal\ for their entire IC~348
sample.  Note that since we are measuring IR color excesses, we are
somewhat less sensitive to errors in $A_K$ than if we were simply
measuring bandpass magnitudes.

As for the spectral types, the Luhman spectral types tend to be
$\approx$1~subclass earlier than those of Najita~\etal\@ This is
expected, given the spectral type calibration chosen by Najita~\etal\
(see their Figure 7, bottom panel).  If the Luhman (1999) types are
adopted, the net result is that the IR excesses we measure would be
larger, strengthening our finding that IR excesses are common in this
sample.  Of course, the mass estimates would be increased, pushing some
objects from brown dwarf to VLM stars, but this would not change our
findings about the high frequency of IR excesses in this very low mass
sample.  Likewise, our analysis of the inner hole sizes
(\S~\ref{sec:holes}) would not be impacted much, since the maximum
\EKmLp\ for the models has only a weak dependence on spectral type for
M5 to M9 objects (Figure~\ref{plot-kl-holes}). Hence, the uncertainties
in the spectral types and extinctions do not seriously affect our
analysis.

\subsection{Contamination from Disk Emission \label{sec:contamination}}

Disk emission can in principal lead to systematic errors in the derived
spectral types and extinctions.  This is unlikely to be an issue for the
Taurus and IC~348 outer samples, since the spectral typing was done at
optical wavelengths ($I$-band) where any disk emission should be
negligible.  However, disk emission contamination is potentially more
important for the IC~348 core sample, whose spectral types are based on
NICMOS 1.6$-$2.2~\micron\ narrow-band photometry measuring the
1.9~\micron\ \htwoo\ absorption band.  In a standard passive and/or
accretion disk where the flux excess follows the form $\Delta
F_{\lambda} \propto \lambda^{-4/3}$, disk emission would lead to weaker
\htwoo\ absorption, and hence earlier spectral types.

We use our passive disk models to quantify the impact of disk emission
on the NICMOS spectral types.  We explore the worst-case scenario, in
which the disks do not have inner holes and hence the disk emission is
maximized.  Our approach is as follows: for each object, we construct
passive disk models based on the measured spectral type and spanning a
wide range of viewing angle $\theta$.  We use the observed \EKmLp\ to
assign a corresponding value of $\theta$ and then determine the amount
of disk excess in the NICMOS narrow-band filters.  Employing the
Najita~\etal\ calibration of NICMOS photometry to spectral type, we then
determine the change in typing.  (For a fully consistent analysis, we
would do this iteratively, using the inferred error in spectral type to
recompute the observed \EKmLp, and then repeating the above process to
re-determine the systematic error in spectral type.  In practice, this
is unnecessary since the disk contamination is small.)

For the IC~348 core sample, we find the median effect of contamination
by disk emission is 0.2~subclasses, \ie, the objects should be
classified as slightly later types. The median excess in the $K$-band
photometry is 0.07~mags.  As expected, the objects with the most
significant disk contamination are those with the earliest spectral
types and largest \EKmLp: here, the effect of disk emission can be as
large as $\approx0.5-0.7$~subclasses, producing a $K$-band excess of
$\approx0.2-0.3$~mags.  Even in these cases, the consequences are minor:
the effect of a later spectral type is to reduce the measured \EKmLp.
However, the actual impact is small since the photospheric \KmLp\ color
changes slowly with spectral type for late M-dwarfs.  Therefore, we
conclude that the contamination from disk emission is not significant.
In a similar vein, the impact of disk emission on the extinctions
derived by \citet{2000ApJ...541..977N} is small.

Again, since these calculations use models without inner holes, the
above results present the worst case scenario for disk contamination.
In the event that the disks have inner holes of radii
$\gtrsim2R_{\star}$, as we infer in \S~\ref{sec:holes}, the effect of
disk emission on the NICMOS-derived spectral types and extinctions is
negligible.

\subsection{Surface Gravity \label{sec:gravity}}

We have used the colors of field M~dwarfs in order to identify which of
our targets have intrinsic IR excesses.  However, our targets are
pre-main sequence objects, and hence will have lower surface gravities
than the field M~dwarfs.  The sense of this effect is likely that at
objects with lower gravities have bluer \KmLp\ colors.  This is seen in
Figure~\ref{plot-passive-2panel}, where the \KmLp\ colors of giant stars
(spectral types G0~III to M5~III) are comparable to or bluer than dwarf
stars (A0~V to M9.5~V).  Pre-main sequence stars are intermediate in
surface gravity, and hence we expect their \KmLp\ colors to be no redder
than the dwarf stars.  If they are bluer than dwarfs, then we would
infer that the IR excesses are larger and even more common.

Given the current uncertainties in the theoretical atmospheres, we chose
not to use the models to measure \EKmLp\ in \S~3, but instead used the
empirical field dwarf locus.  Here, we examine the
\citet{1998A&A...337..403B, 2002A&A...382..563B} models to explore the
{\em relative} shift in color due to gravity
effects. Figure~\ref{plot-gravity} plots models with ages of 1~Myr and
1~Gyr: there is a difference of a factor of $\approx$30 in the surface
gravities at fixed mass. The \Teff\ for the older (1~Gyr) models have
been converted to spectral type using results for field M~dwarfs from
\citet{2000ApJ...535..965L, 2001ApJ...548..908L} and
\citet{1996A&A...305L...1T}, fitted to a linear relation:
\begin{equation}
\Teff (K) = 3851 - 185.3 \times SpT, \label{eqn:dwarfscale}
\end{equation}
where $SpT$ is the M~spectral subclass.  For the younger (1~Myr) models
we use either this scale or the aforementioned
\citet{1999ApJ...525..466L} scale (equation~(\ref{eqn:luhmanscale})).
Both the ``clear'' and the ``dusty'' atmospheric models predict that
younger, lower gravity objects have bluer colors at fixed spectral type.
The amplitude of the difference varies with the choice of model and
temperature scale, but generally lower gravity objects are predicted to
be 0.1~mag bluer in \KmLp\ at fixed spectral type.

\begin{figure}
\vskip -0.5in
\hskip -0.5in
\centerline{\includegraphics[width=4.25in,angle=90]{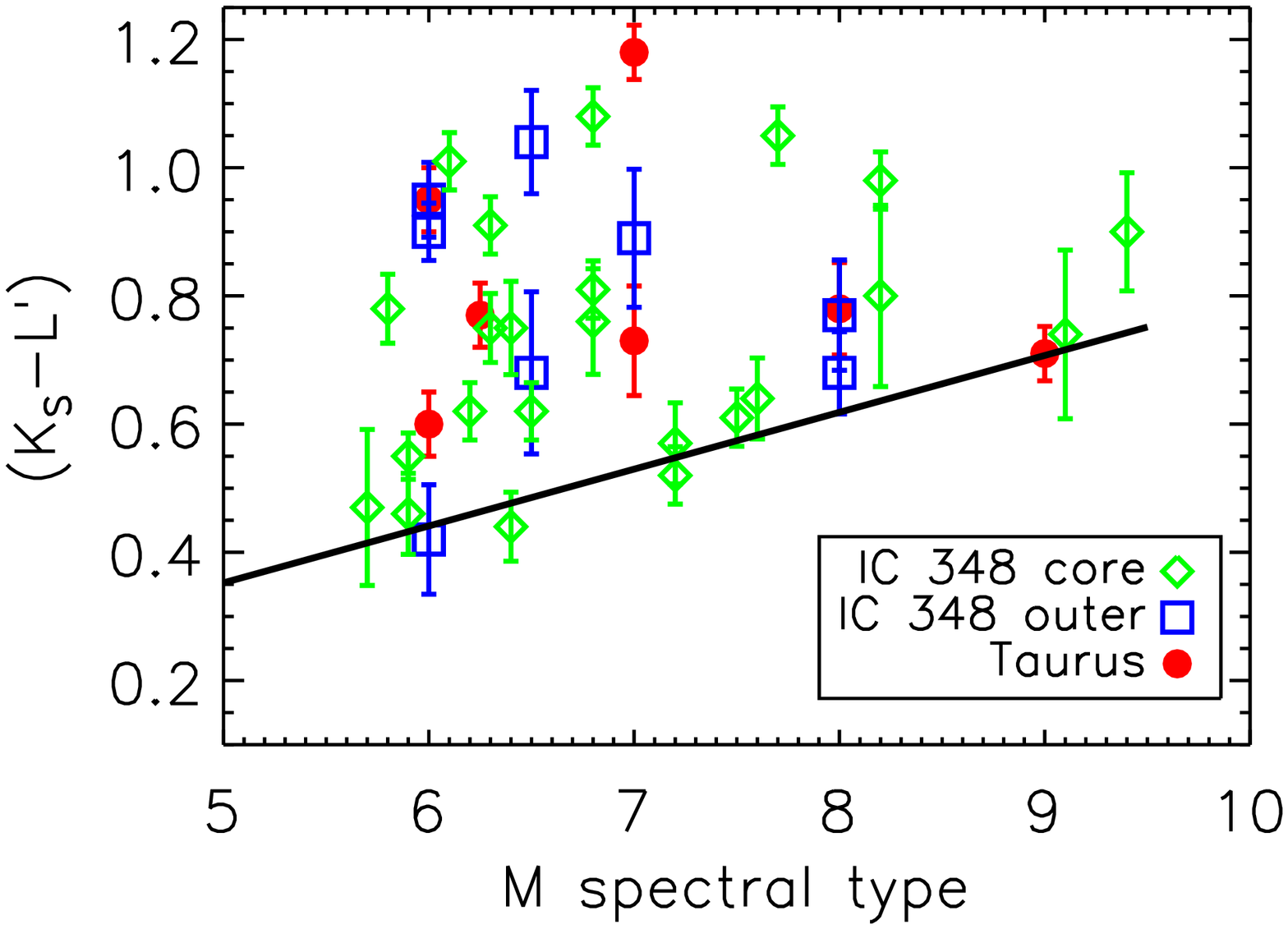}}
\vskip -0.25in
\hskip -0.5in
\centerline{\includegraphics[width=4.25in,angle=90]{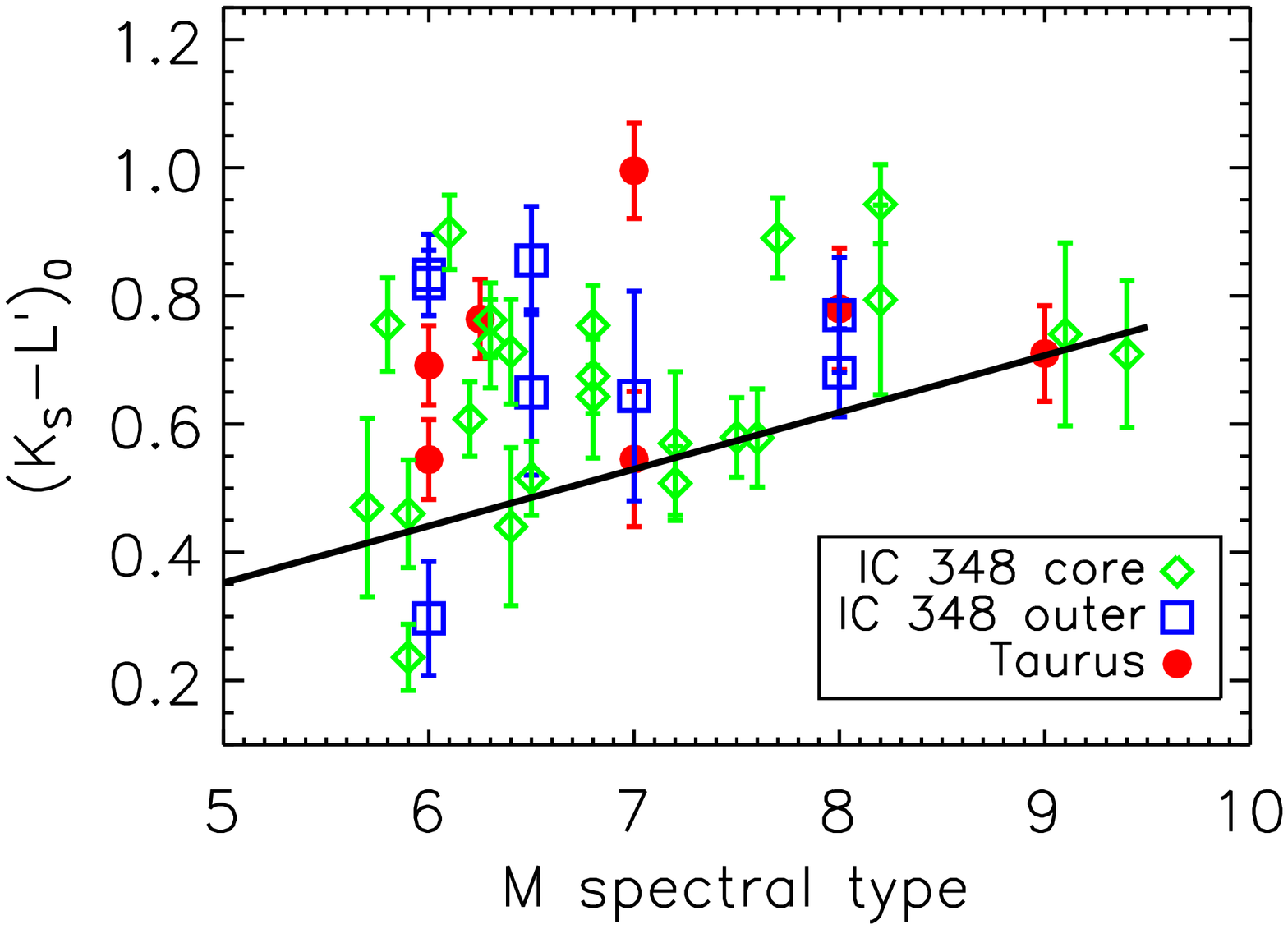}}
\vskip -2ex
\caption{\normalsize {\bf Top:} Observed \KmLp\ colors as a function of
spectral type.  Typical errors in the spectral type are 0.5 subclasses
or less.  The heavy line represents the colors of field M~dwarfs (see
\S~\ref{sec:2mass} for references).  {\bf Bottom:} Dereddened \KmLp\
colors.  Most of the objects show IR emission in excess of that expected
from their photospheres. \label{plot-kl}}
\end{figure}

\begin{figure}
\hskip -0.25in
\centerline{\includegraphics[width=5in,angle=90]{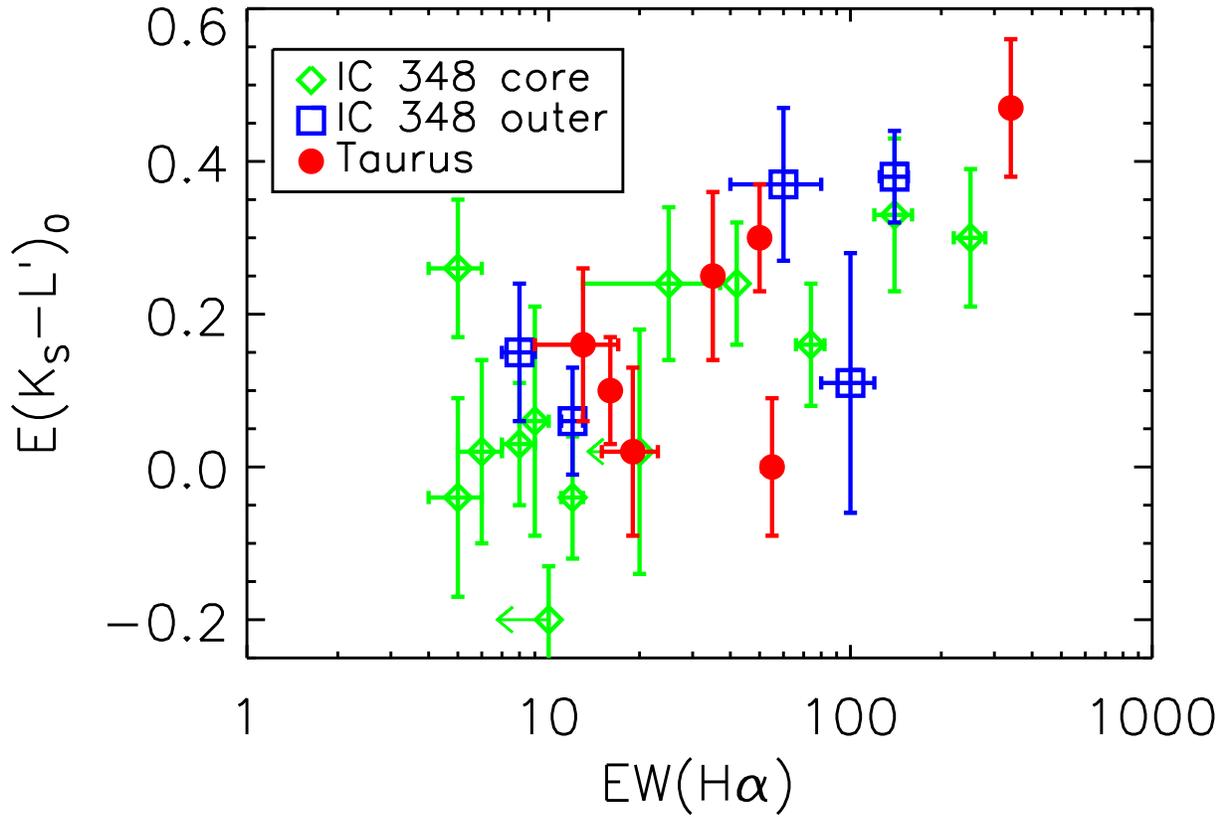}}
\caption{\normalsize Relation between intrinsic IR excess and equivalent
width of H$\alpha$ emission in Angstroms for the 23~objects where such
data are available along with H$\alpha$ upper limits for 2~objects.  The
two quantities are correlated at the 3$\sigma$ level, based on a
Spearman rank correlation test.  This supports the idea that the optical
and near-IR emission both originate from the same phenomenon, namely
circumstellar accretion disks. \label{halpha}}
\end{figure}

\begin{figure}
\vskip -1in
\hskip -0.25in
\centerline{\includegraphics[width=4.5in,angle=90]{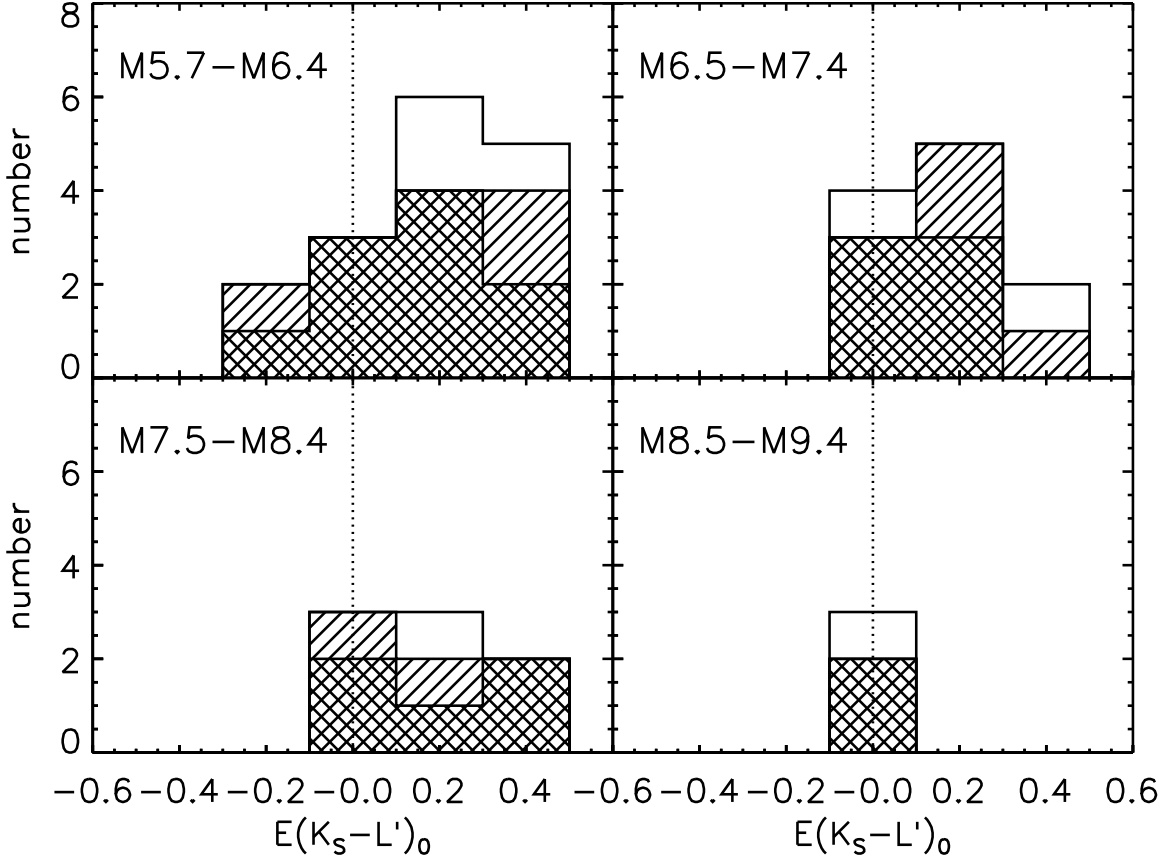}}
\vskip 4ex
\caption{\normalsize Histogram of \KmLpo\ excesses as a function of
spectral type. The median $\pm1\sigma$ measurement errors
($\pm$0.09~mag) are comparable to the bin width. The clear bins
represent the Taurus sample; the single-hatched bins represent the outer
IC~348 sample \citep{1999ApJ...525..466L}; and the double-hatched
bins represent the IC~348 core sample \citep{2000ApJ...541..977N}.
The distribution of excesses among the three earliest spectral type bins
(M5.7--M6.4, M6.5--M7.4, and M7.5--8.4) are consistent with being drawn
from the same parent population based on the K-S test.  Hence, there is
no strong evidence for the IR excesses being dependent on the mass of
the central object.  \label{klexcess-hist}}
\end{figure}

\begin{figure}
\vskip -1in
\centerline{\includegraphics[width=4.5in,angle=90]{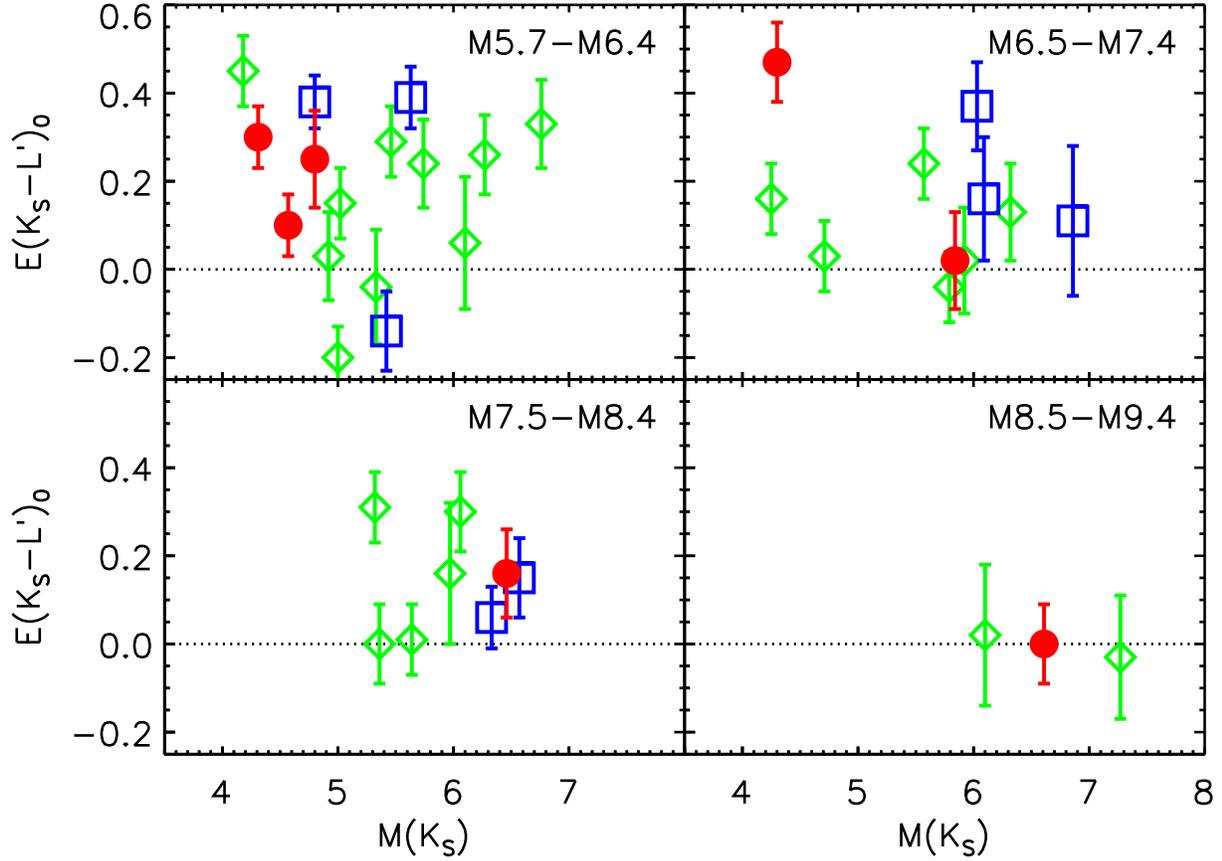}}
\vskip 4ex
\caption{\normalsize IR excess as a function of dereddened absolute
\Ks-band magnitude.  At fixed spectral type, \MKs\ is a proxy for the
relative age, since the evolutionary tracks for young objects are
approximately constant in temperature for a given mass.  No
statistically significant correlation with \MKs\ exists for each
spectral type bin, suggesting the inner disk regions do not evolve
substantially over the first $\sim$3~Myr. The symbols represent the
different sub-samples and have the same meanings as in
Figure~\ref{plot-hrd}. Errors in \MKs\ are $\approx0.2$~mag, about twice
the size of the plotting symbols. \label{agetrend}}
\end{figure}

\begin{figure}
\hskip -0.25in
\vskip -0.5in
\centerline{\includegraphics[width=5in,angle=0]{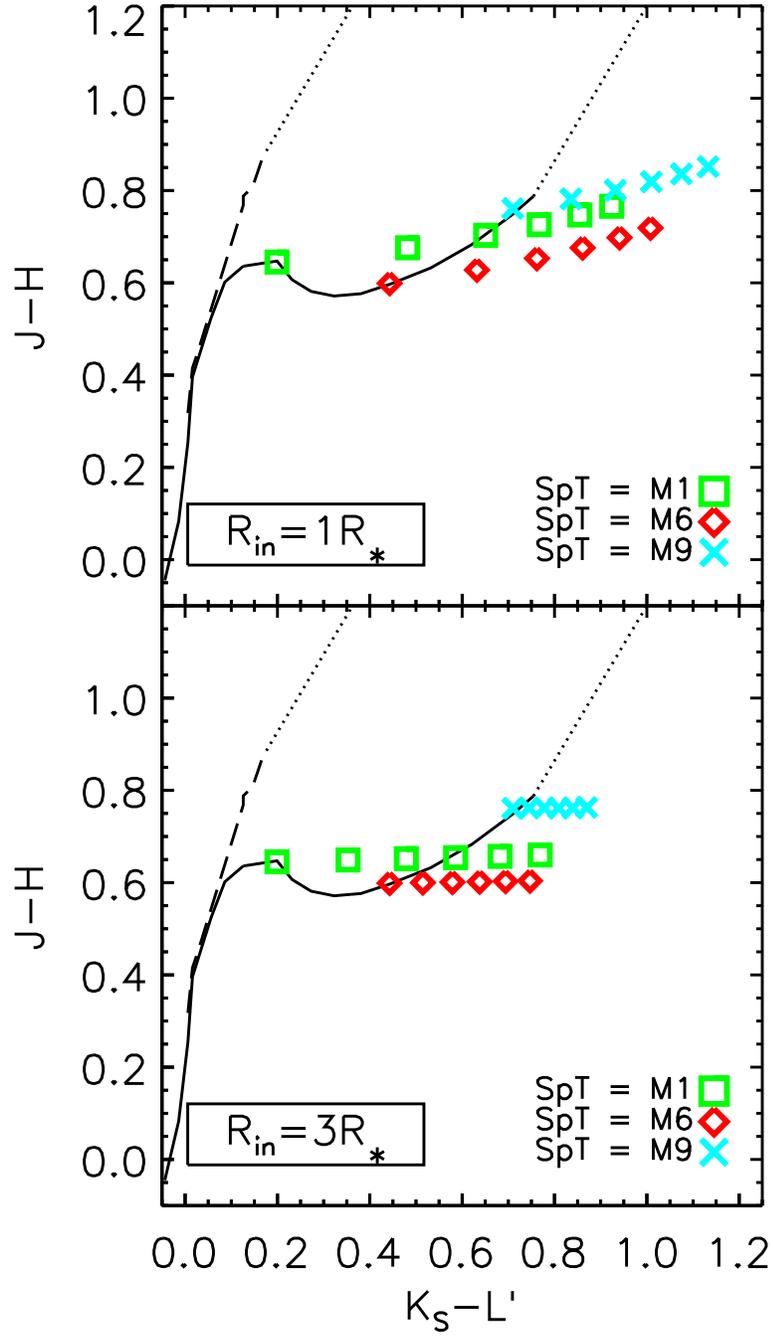}}
\vskip 5ex
\caption{\normalsize Flat blackbody reprocessing disk models plotted on
\JHKsLp\ color-color diagrams.  The locus of main sequence stars is
shown as a solid line and that of giant stars as a dashed line. (See
\S~\ref{sec:models} for details.)  The two dotted lines represent the
reddening vector, and the area in between is the locus of reddened
stars.  Each plot shows models with different inner disk radii, and the
symbols indicate the spectral type of the central star.  The symbols are
equally spaced in $\cos\theta$ from 0 to 1, where $\theta$ is the
viewing angle.
\label{plot-passive-2panel}}
\end{figure}

\begin{figure}
\hskip -0.25in
\centerline{\includegraphics[width=5in,angle=90]{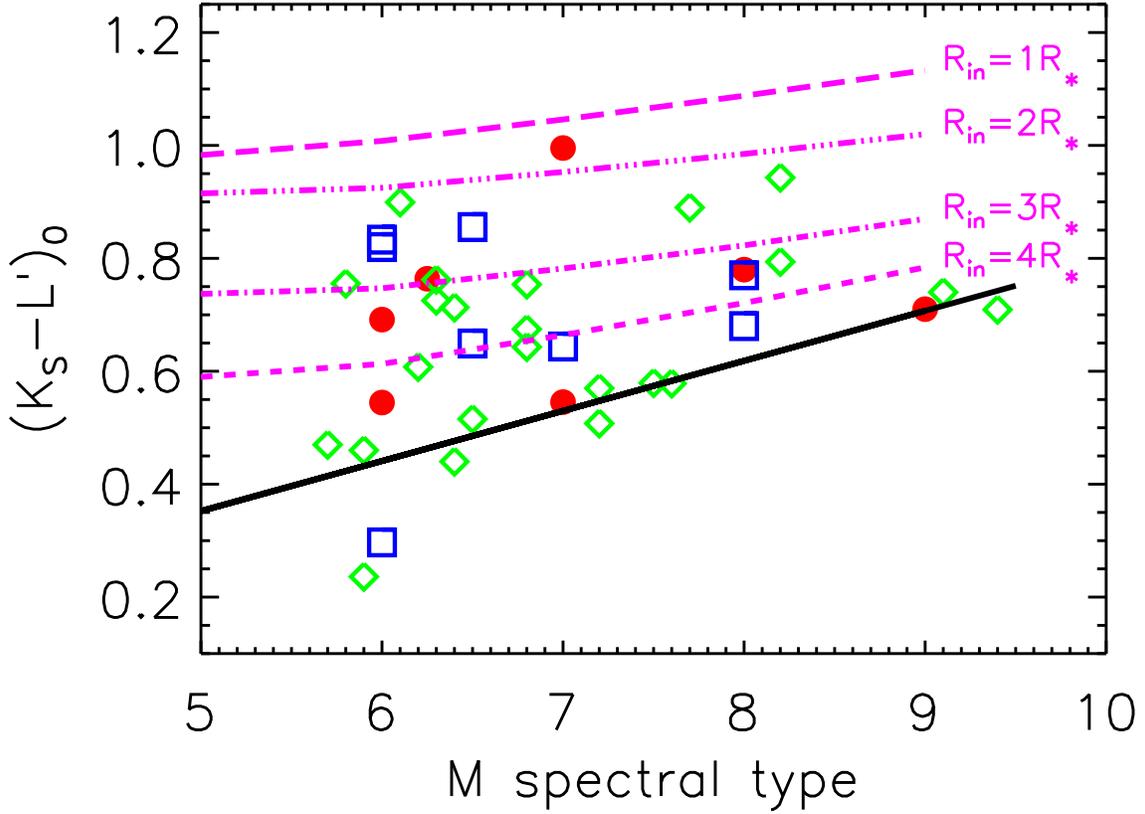}}
\caption{\normalsize Dereddened \KmLp\ colors for our targets (same data
and symbols as in Figure~\ref{plot-kl}) compared with face-on
reprocessing disk models.  Disk models with different inner hole radii
are labeled in units of the stellar radius, \ie, $R_{in}=1\Rstar$ means
there is no hole.  The maximum observed IR excesses are less those
expected from a disk with no inner hole, and suggests
$R_{in}\gtrsim2\Rstar$ is the more likely situation.  Notice also
that the observed small/non-existent \KmLpo\ excesses of the coolest
objects, which are spectral type M9, suggest either disks with very
large inner holes or no disks at all. \label{plot-kl-holes}}
\end{figure}

\begin{figure}
\centerline{
\includegraphics[width=2.6in,angle=0]{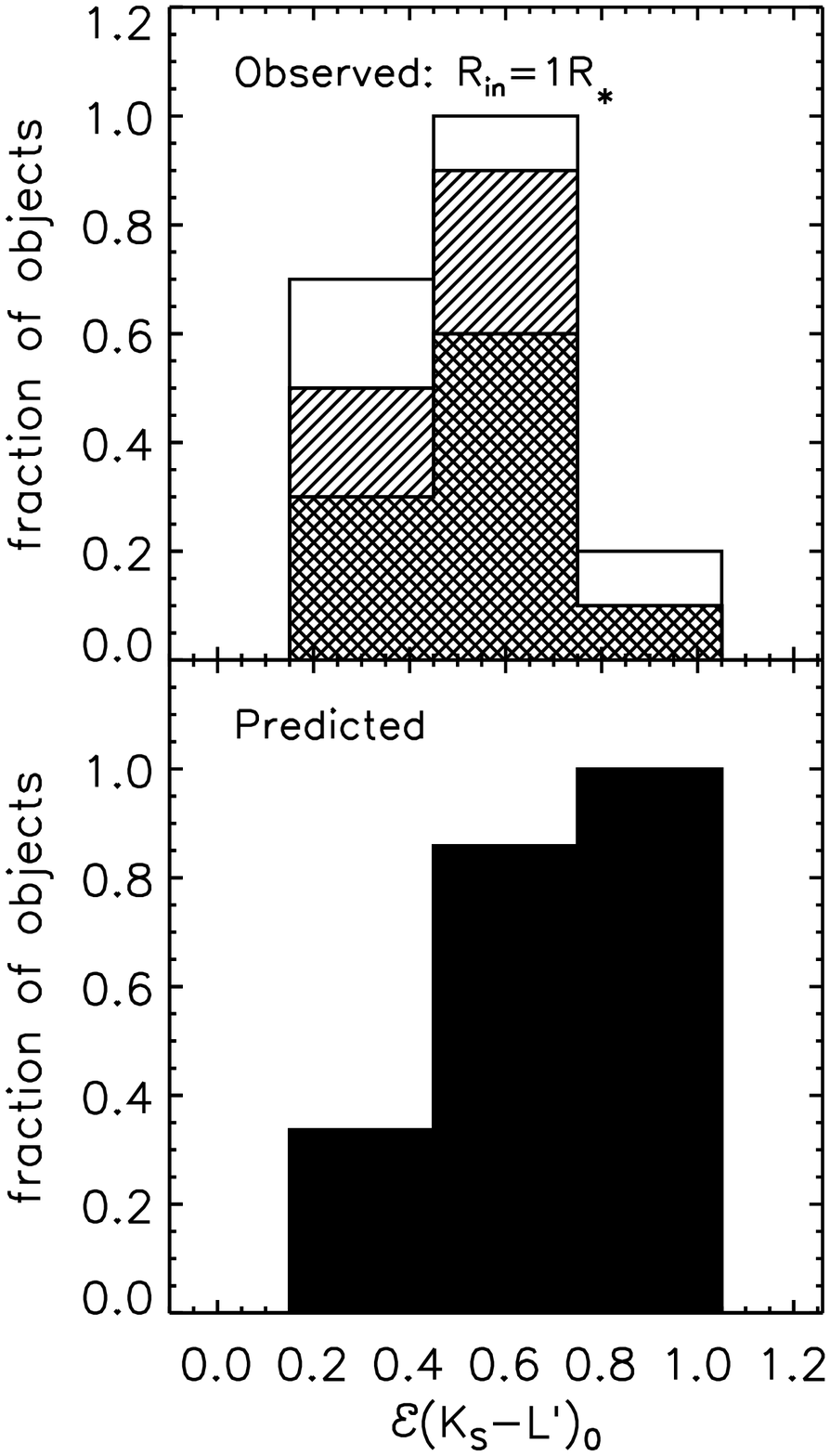}
\hskip -0.4in
\includegraphics[width=2.6in,angle=0]{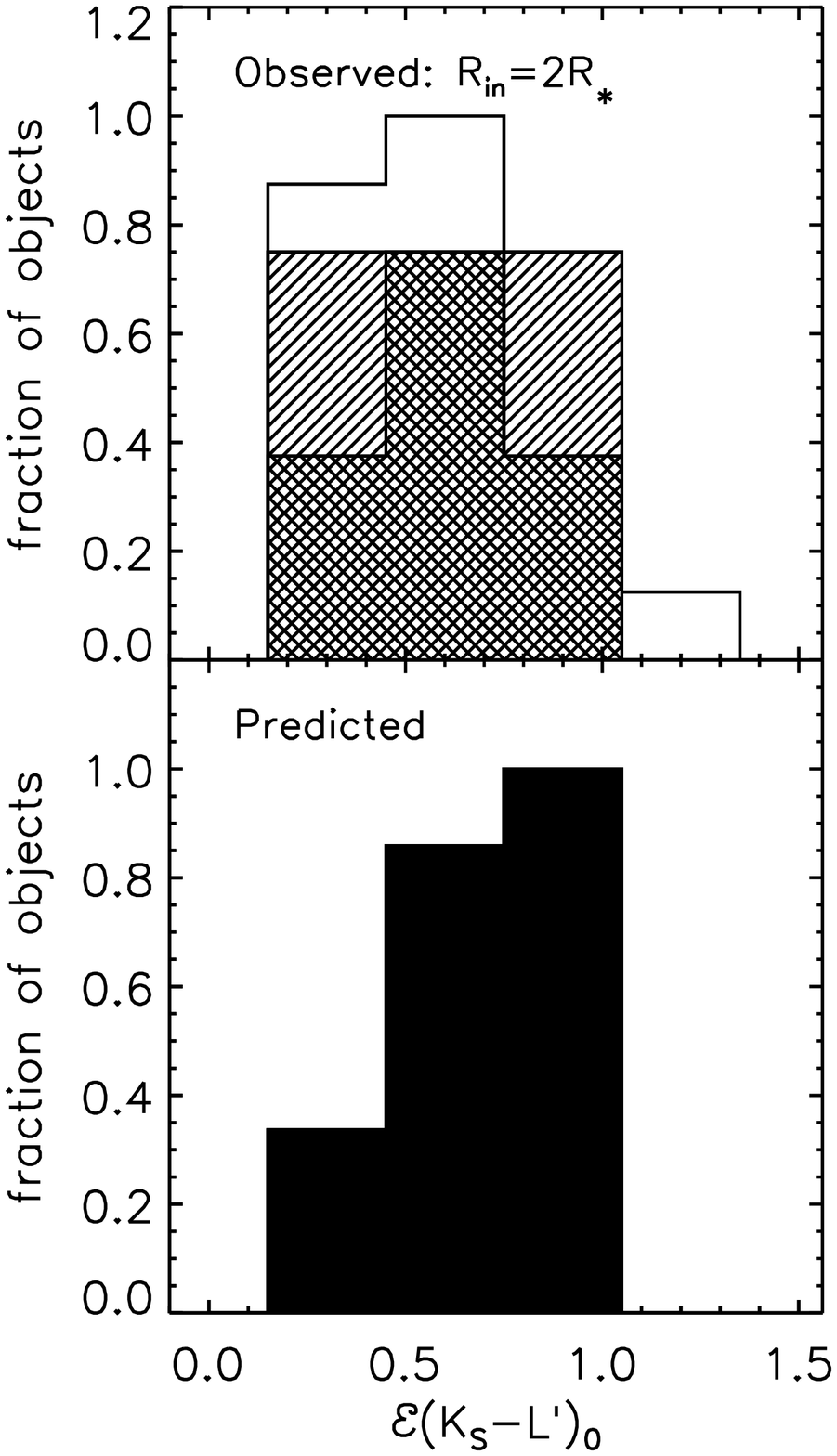}
\hskip -0.4in
\includegraphics[width=2.6in,angle=0]{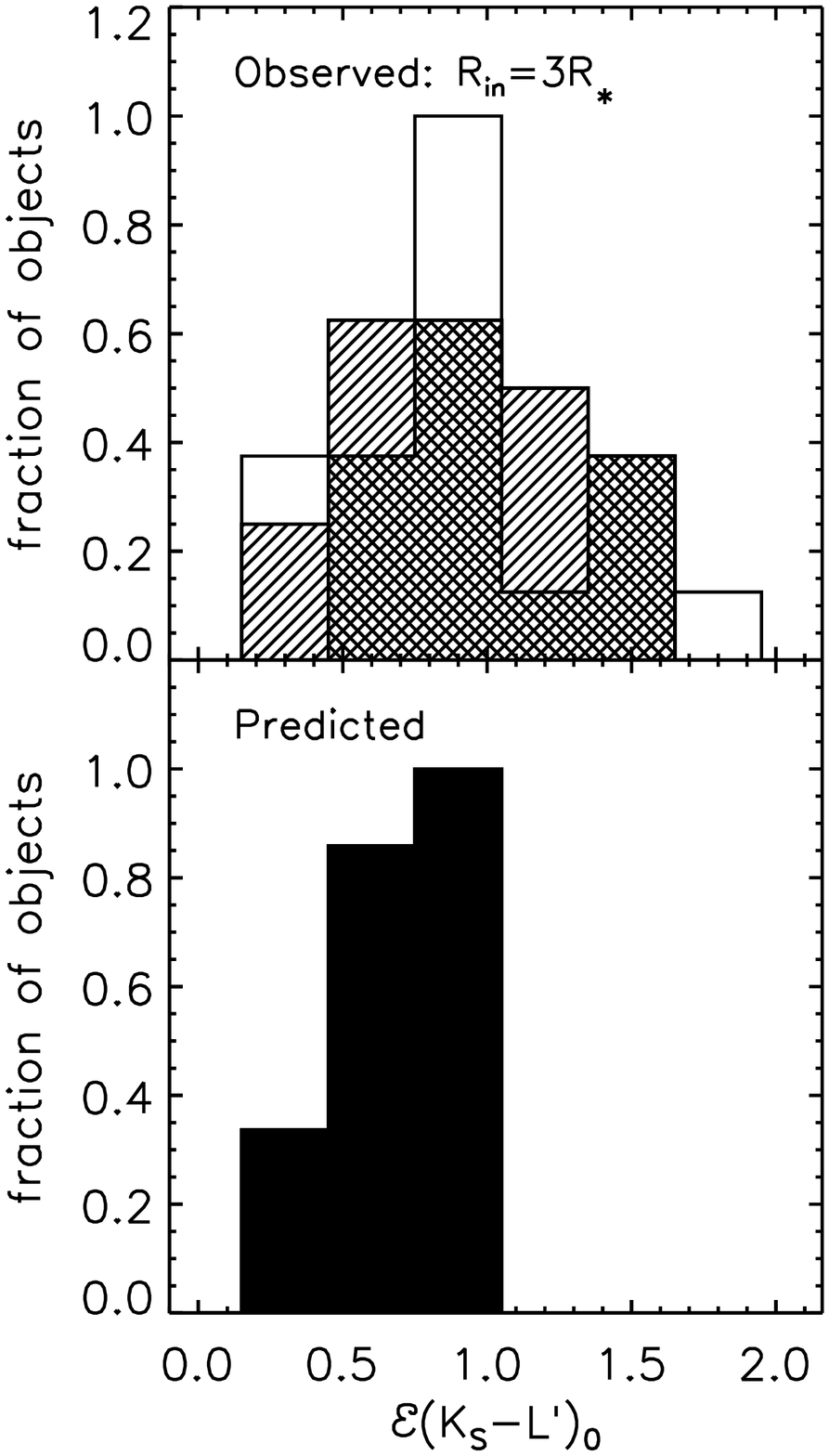}
}
\vskip 4ex
\caption{\normalsize Comparison of the observed IR excess distribution
with passive disk models possessing different inner radii.  Disk models
with inner radii ($R_{in}$) of 1, 2, and 3\Rstar\ are shown.  For a
given disk model, the observed \KmLpo\ excess is divided by the
maximum possible model excess (\ie, a face-on disk) to form the
normalized color excess \cEKmLp, which is independent of the central
object's temperature.  The top panels show the observations. Different
shadings indicate different sub-samples: Taurus ({\em clear bins}),
IC~348 outer sample ({\em single-hatched bins}), and IC~348 core sample
({\em double-hatched bins}).  The bottom panels plot the normalized
color excess distribution for a set of disk models seen from randomly
chosen viewing angles.  The observations agree best with the
$R_{in}=2\Rstar$ models, suggesting that disks around young brown dwarfs
have inner holes.  See \S~\ref{sec:holes} for details.
\label{holes}}
\end{figure}

\begin{figure}
\vskip -0.5in
\centerline{\includegraphics[width=4in,angle=90]{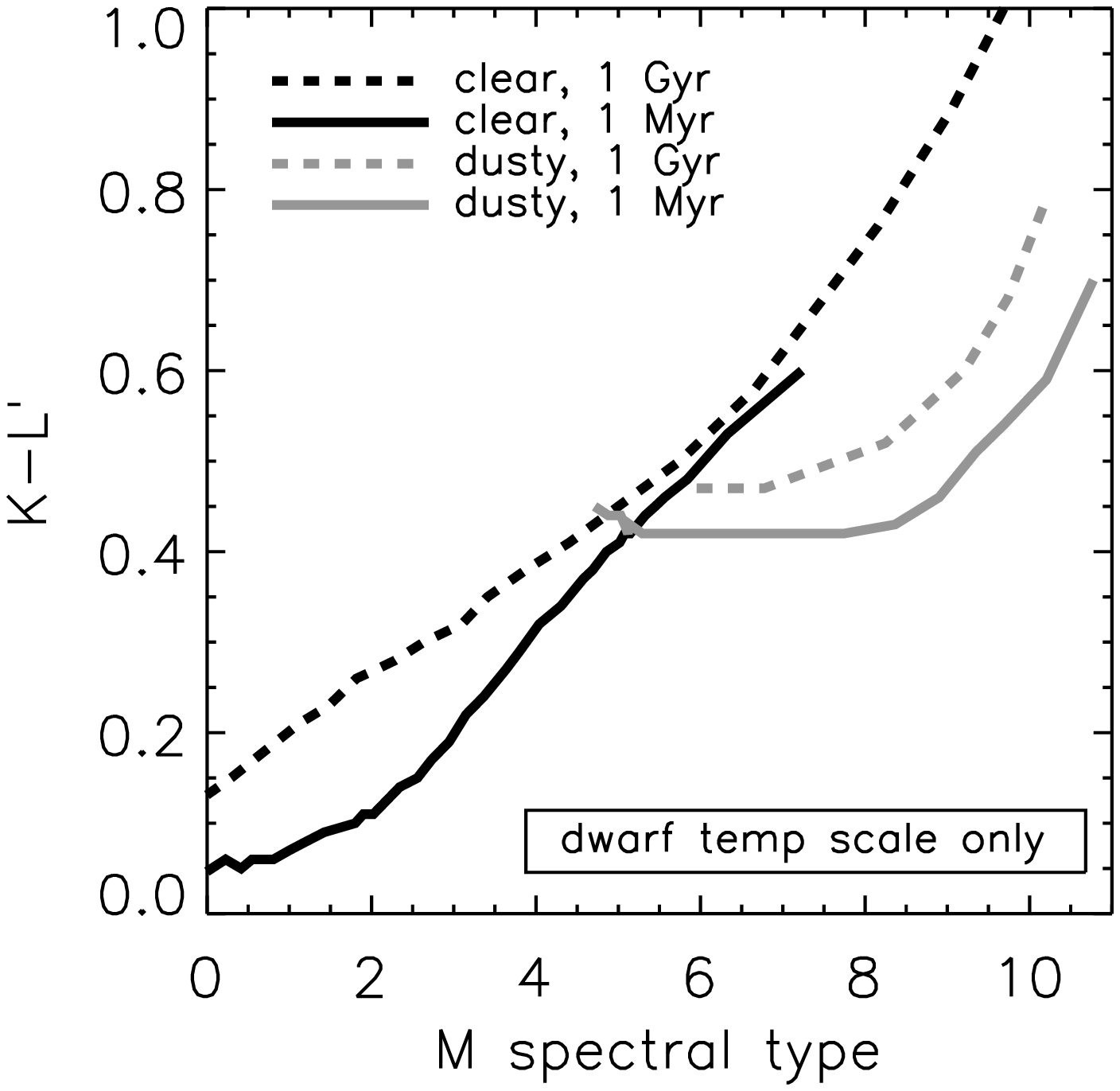}}
\vskip -0.25in
\centerline{\includegraphics[width=4in,angle=90]{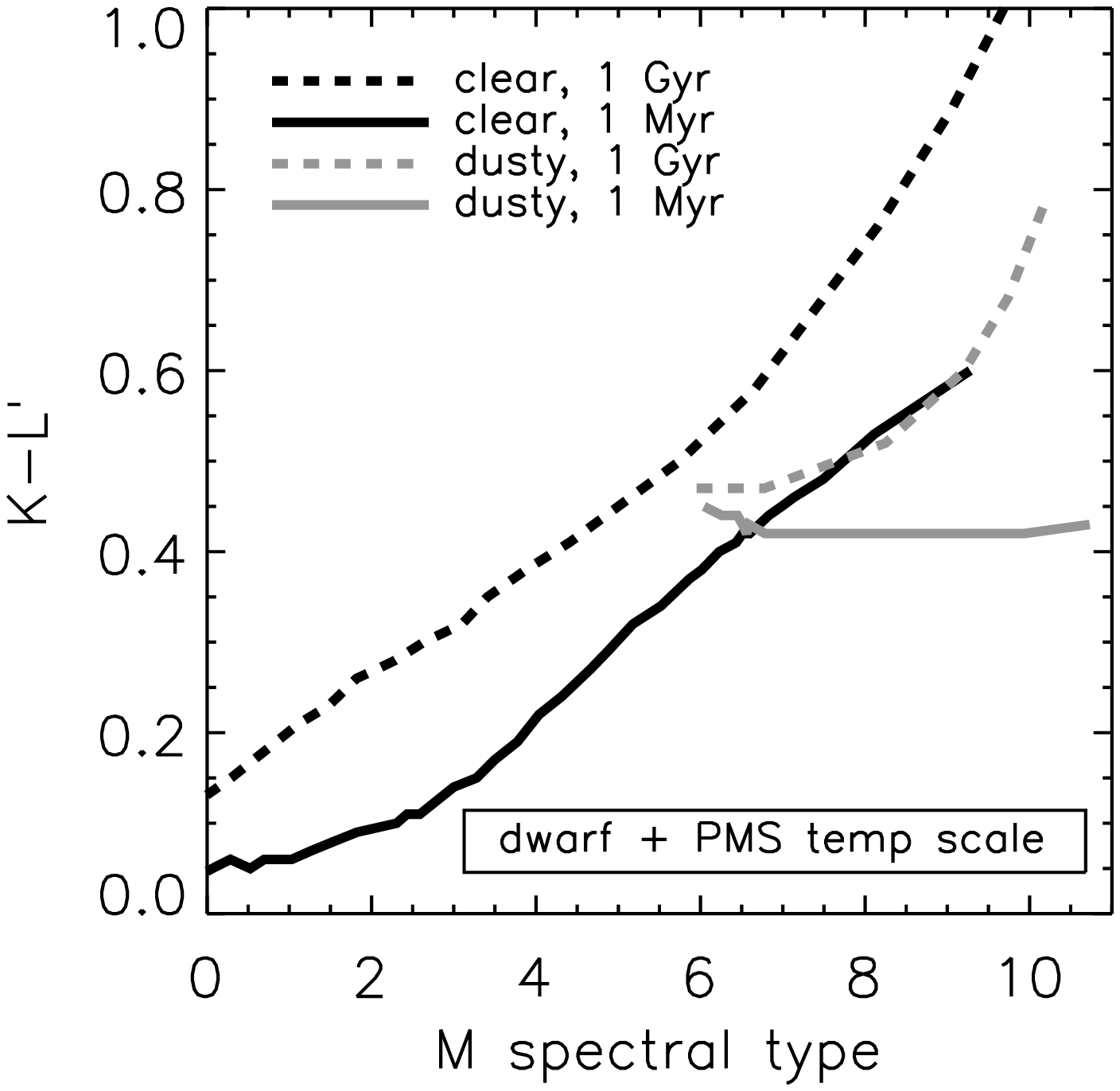}}
\caption{\normalsize The effect of surface gravity on $K-L^\prime$
colors, based on models by \citet{1998A&A...337..403B,
2002A&A...382..563B}.  Dust-free (``clear'') models are plotted in
black, and dusty models in grey.  Models with ages of 1~Gyr (higher
surface gravity) are converted to spectral types using a dwarf
temperature scale. Models with ages of 1~Myr (lower surface gravity) use
either the dwarf temperature scale ({\em upper plot}) or the Luhman
(1999) pre-main sequence scale ({\em lower plot}). For our objects'
spectral types, the lower surface gravity of young objects is predicted
to lead to $\approx$0.1~mag bluer colors; the observations suggest the
effect is even smaller.  See \S~\ref{sec:gravity} for details.
\label{plot-gravity}}
\end{figure}

Perhaps the most convincing evidence that gravity effects do not
significantly affect the photospheric colors can be seen in the data
themselves.  The loci of the observed and dereddened \KmLp\ colors in
Figure~\ref{plot-kl} both show a lower envelope which agrees well with
the empirical dwarf locus, to better than $\lesssim0.05$~mag.  This
level of difference has little bearing on our analysis.\footnote{One
interesting feature is that the dusty models predict fairly constant
$K-\Lp$ colors for objects with mid to late-M type objects, especially
when using the \citet{1999ApJ...525..466L} temperature scale.  This is
inconsistent with the observations.  For field M6$-$M9 dwarfs, the
models are $\approx0.1-0.2$~mag bluer than the observed colors (\eg, the
solid line plotted in Figure~\ref{plot-kl}). Likewise, the dusty models
have bluer \KmLp\ colors than the observed lower envelope for our sample
of young M~dwarfs (Figure~\ref{plot-kl}).  However, our sample contains
only a few young M9 objects, so the observations would allow for
constant $K-\Lp$ colors if this effect in fact begins around M8, instead
of at M6 as currently predicted by the dusty models. If this is the
case, then the M9 objects would have actually small IR excesses, whereas
they are currently inferred to have almost none.}

\subsection{Sample Selection Biases}

Biases in the sample selection could impact our conclusions. For
example, if brown dwarfs without disks are systematically excluded from
our sample, then our estimate of the disk-bearing fraction will be
incorrect. The effect of extinction on the sample selection is probably
the primary concern.  Objects with edge-on disks will be heavily
extincted and are almost certainly missing from the optical/IR imaging
surveys used to construct our sample.  (Indeed, this bias is also
present in most previous T~Tauri disk studies using broad-band
photometry.) Viewing angles of greater than $80\degs$, which is the
cutoff value we adopt in \S~\ref{sec:holes} for obscured edge-on disks,
account for 17\% of a random distribution of disk inclinations, so the
disk fraction for our sample would be underestimated by only a small
amount ($\approx$3\%).  Also, including such objects would only act to
increase the measured disk fraction, which we already find is very high.

One common approach to minimizing selection biases is to use a complete
sample, where all the objects in a given area on the sky are identified
to a well-defined limit.  The \citet{2000ApJ...541..977N} survey is the
source of our IC~348 core sample; this survey is $K$-band selected and
has a well-defined completeness limit for spectral classification.  We
have \Lp-band observations for nearly every cluster member from their
survey in the M5.7 to M9.4 range; only 2~objects out of 26 were not
observed.  Also, most of our targets are several magnitudes brighter in
$K$-band than the survey limit.  Therefore, we expect that our IC~348
core sample is essentially complete, and that selection biases due to
extinction are negligible.  The trends we identified in \S~3 for our
entire \Lp-band sample are also found if we consider only the IC~348
core sample.  This suggests that selection biases are probably not a
significant effect when analyzing the \Lp-band sample as a whole.

The three coolest/lowest mass objects, all with spectral type
$\approx$M9, have negligible IR excesses.  This might be due to a
selection effect if objects with larger excesses tended to be
preferentially extincted and hence missing from the sample.  But there
is no evidence for this effect: for the M5.7 to M7.7 objects, no
statistically significant ($\gtrsim1\sigma$) correlation exists between
\EKmLp\ and $A_K$.  Also, even the M9 objects are 1.5~mags above the
$K$-band survey limit of Najita~\etal\@ Thus, it is unlikely there are
M9 objects with large excesses which are missing from our sample because
of large extinctions.  However, a much larger sample of M9 (and cooler)
sources are needed to accrurately assess the disk fraction of such low
mass objects ($\lesssim15-20$~\Mjup).


\subsection{Pre-Main Sequence Temperature Scale and Mass Estimates
   \label{sec:tempscale}} 

When analyzing the trends in the disk properties (\S~3), we used the
observable quantities of spectral type and \MKs\ as surrogates for the
relative masses and ages of the targets.  Preferably, one would like to
use absolute values for masses and ages, as obtained from theoretical
models.  However, an examination of Figure~\ref{plot-hrd} shows the
problem: different models give very different results.  Also, the
current models do not always cover the observed location of the young
brown dwarfs in the plot.  (This is true using either $K$-band absolute
magnitude or bolometric magnitude as the ordinate.)

An important caveat is the choice of the conversion between \Teff\ and
spectral type.  We have used the \citet{1999ApJ...525..466L} scale
throughout.  If we had used a scale appropriate for field M~dwarfs
(equation~(\ref{eqn:dwarfscale})) then evolutionary models of a given
temperature would correspond to earlier spectral types --- the net
result would be much lower mass estimates for our targets, with masses
of $\lesssim$10~\Mjup\ for the coolest ones.  Hence our choice of the
\citet{1999ApJ...525..466L} scale is a conservative one.  Similarly, the
disk modeling results in Table~\ref{table-passive} would change by
$\le0.05$~mag with a dwarf temperature scale, and our other analyses
using disk models in \S~4 would also be unaffected.

We reiterate that our targets are selected based on their spectral
types.  The resulting sample is significantly cooler than previous
$L$-band surveys, which were sensitive down to spectral types of
$\sim$M5 \citep{1995ApJS..101..117K, 2001AJ....121.2065H}, and hence
comprise the lowest mass sample of young objects studied to date at
these wavelengths, independent of the choice of temperature scale.


\section{Discussion}

\subsection{Disk Frequency of Young Stars and Brown Dwarfs
\label{sec:findingdisks}} 

In our sample of brown dwarfs and VLM stars, we find 31~out of~38
objects possess intrinsic IR excesses ($82\%\pm15\%$, where we quote
only Poisson statistical errors in this section in order to compare with
previously published results).  This disk fraction is comparable to, or
exceeds, the disk fraction for the higher mass stars in the same
star-forming regions.  Using multi-band photometry and spectral types,
\citet{1989AJ.....97.1451S} found 47 out of 83 ($57\%\pm8\%$) T~Tauri
stars in Taurus have significant $K$-band excesses (${\Delta}K \ge
0.1$~mag). (See also \citealp[][]{1995ApJS..101..117K}.)
\citet{2001AJ....121.2065H} found a similar disk fraction of
$65\%\pm8\%$ for stars in IC~348.  
Some caution is in order when comparing our IC~348 results with those of
Haisch \etal\@ The latter used only multi-band photometry without
spectral type information, and hence, as we discuss below, their disk
fraction may be somewhat underestimated.

The most common method for measuring the disk fraction of young stars
uses color-color diagrams based on \JHK, or preferably \JHKL, colors.
Objects with disks are identified as those having IR colors distinct
from reddened dwarf and giant stars.  This method has been used
extensively to study the frequency, properties, and evolution of
circumstellar disks around T~Tauri stars
\citep[e.g.][]{1992ApJ...393..278L, 1995ApJS..101..117K,
1995AJ....109.1682L, 1997AJ....114..288M, 2001ApJ...553L.153H}.  The
method is appealing since only photometry is used, without need for more
time-consuming spectroscopy.  Moreover, the method has been demonstrated
to be effective for T~Tauri stars, \eg, \citet{1998AJ....116.1816H} and
\citet{2001AJ....121.1512H} show that most TTS in Taurus with
multi-wavelength evidence for disks can be identified from \JHK\ colors
alone, and all can be found with \JHKL\ colors.  Here we consider the
value of this method for studying disks around young brown dwarfs.

\begin{figure}
\vskip -1in
\hskip 0.25in
\centerline{\includegraphics[width=6in,angle=90]{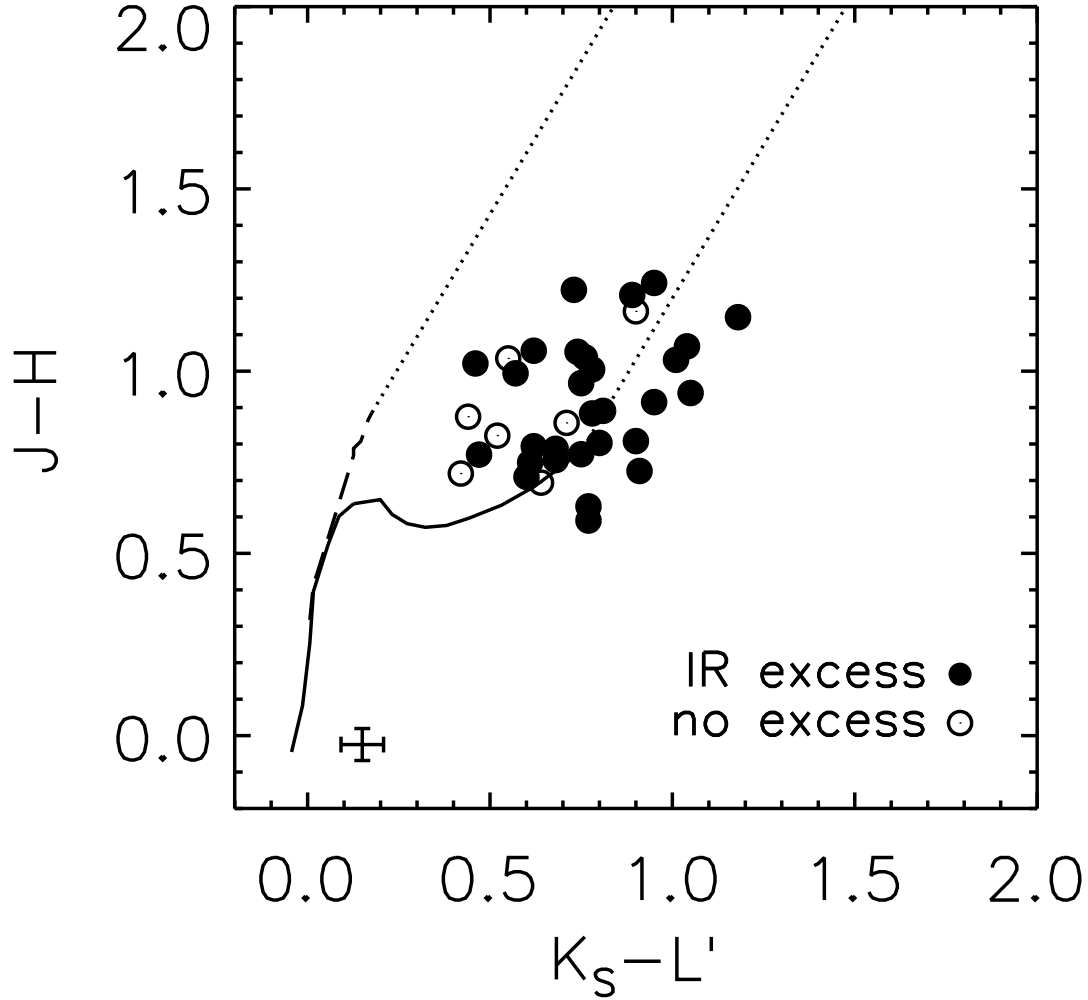}}
\vskip 2ex
\caption{\normalsize \JHKsLp\ color-color diagram for our sample.  The
various lines represent the loci of reddened giant and dwarf stars (see
Figure~\ref{plot-passive-2panel} caption).  Median errors are plotted in
the lower right. Based on a conventional color-color analysis, one would
identify only 11~out of 36~objects (those to the right of the dotted
reddening band) as having IR excesses.  However, our analysis
incorporating the objects' spectral types shows that in fact many more
(31~out of~36) have IR excesses: these are plotted as filled circles
($\bullet$).  The majority of sources with disks are missed in the
color-color diagram because their IR excesses are modest.  A similar
analysis using only \JHKs\ colors would miss nearly all the objects with
IR excesses.
\label{colorcolor}}
\end{figure}

Figure~\ref{colorcolor} illustrates the commonly used IR color-color
analysis applied to our sample.  From this diagram, one would identify
only 11~out of 36~objects as having IR excesses.  (Two objects do not
have 2MASS $J$-band photometry.)  However, our analysis which
incorporates the objects' spectral types and extinctions shows that in
fact many more objects (31~out of~36) have IR excesses. The majority of
sources with disks are missed in the color-color diagram because their
IR excesses are modest.  In fact, an analysis using only \JHKs\ colors,
without \Lp-band data, would find only 2 out of 36 objects with IR
excesses, \ie, essentially missing all the objects with disks.  These
observational results confirm model predictions of very small \JHK\
excesses from brown dwarf disks (e.g., \citealp{2001A&A...376L..22N},
this work).

There are two physical reasons why using only IR colors works poorly.
Both lead to decreased contrast between brown dwarfs and their disks.
(1) Because brown dwarfs are less luminous than the higher mass T~Tauri
stars, disks around brown dwarfs will be cooler and less luminous;
therefore, the corresponding IR excesses will be smaller (\eg,
Table~\ref{table-passive}). (2) T~Tauri stars span a limited range in
photospheric \KmLp\ color, and hence even a modest disk excess will
produce IR colors readily distinguishable in color-color diagrams.  In
contrast, young brown dwarfs, with spectral types of M6 and later, have
redder photospheres and span a larger range in intrinsic \KmLp\ color
(\eg, Figure~\ref{plot-passive-2panel}).  Hence, the modest IR excesses
produced by their disks will be harder to distinguish than for the case
of T~Tauri stars.

This latter item raises the fact that color-color analyses inevitably
underestimate the disk fraction.  Near-IR imaging samples will include
objects with a range of spectral types.  Earlier-type (hotter) stars
with small IR excesses will not be as readily identified as disk-bearing
in color-color diagrams compared to stars with larger excesses or
later-type (cooler) objects which lie close to the reddening boundary.
This effect leads to an underestimate of the disk fraction.  It is
unlikely to be an significant error in surveys of T~Tauri or Herbig
Ae/Be stars, since they have a small range in intrinsic IR colors.  It
will be a more serious effect for young brown dwarfs.

In a similar vein, the disk fractions inferred from color-color analyses
are very sensitive to the assumed boundary of the reddened stellar locus
(the rightmost line in Figure~\ref{colorcolor}).  The boundary is
determined by the colors of the latest spectral type objects. However,
since spectroscopy is generally not available in these studies, this
boundary must be estimated, which can produce either an underestimate or
overestimate of the true disk fraction.  Since nearly all such surveys
are magnitude-limited, this systematic error can be aggravated by the
variable amounts of extinction to different sample members and hence an
ill-defined spectral type boundary.

Our approach of examining a sample of objects with spectral type
determinations is far more sensitive to disk emission than searches
based on $J\!H\!K\!(L)$ photometry alone.  
Indeed, our results suggest that both \Lp-band data and spectral types
are required in order for an accurate ground-based census of disks
around young brown dwarfs.


\subsection{Comparison with Near-IR Surveys of Trapezium and $\sigma$ Ori}

Two recent surveys have addressed the topic of disks around young VLM
stars and brown dwarfs using large samples in Orion.  Here we compare
their findings with our own results.

\citet{2001ApJ...558L..51M} have studied the disk fraction for young
brown dwarfs in the Trapezium cluster using \JHK\ imaging.  They select
a large sample (109~objects) of brown dwarf candidates identified from
their $(J\!-\!H, H)$ color-magnitude diagram in conjunction with a 1~Myr
theoretical model from \citet{1998A&A...337..403B}.  Then they use \JHK\
color-color diagrams to infer a disk fraction of $\sim65\%\pm15\%$.
Many of their objects with IR excesses clearly possess disks, because
they are coincident with ``proplyd'' sources seen in \HST\ optical
images.

Interpreting their results is complicated by the uncertain masses of
their candidates.  Contamination from background stars is predicted to
be low for the Muench \etal\ magnitude limits
\citep[e.g.][]{2000ApJ...540..236H}, so most of the objects probably are
cluster members.  But brown dwarf selection from the IR color-magnitude
diagram is clearly not robust: about 15\% of their candidates have \JHK\
colors consistent with reddened stars of spectral types M0 or earlier.
Moreover, there is a puzzling inconsistency in the photometric selection
when compared to the available spectroscopy. Ten sources in the Muench
\etal\ survey have spectral types of M6 or later as measured by
\citet{1997AJ....113.1733H}, indicating that they are near or below the
substellar limit. However, all of these are brighter than the 1~Myr
model predictions for a 0.08~\Msun\ object, some by several
magnitudes. The evolutionary models would thus indicate that these
objects have stellar masses based on their brightnesses and are not
brown dwarfs.  Finally, the Muench \etal\ selection approach essentially
uses unextincted $J$-band magnitudes, and therefore depends on the
assumed age of the Trapezium cluster.  Objects older than 1~Myr, and
hence higher mass, might be included in the sample.  For example, for an
assumed age of 3~Myr, the brightest candidates would correspond to a
mass of 0.15~\Msun\ based on the \citet{1998A&A...337..403B} models.
Spectroscopic follow-up will be important for refining the mass
estimates of their brown dwarf candidates.

The location of the Muench \etal\ Trapezium brown dwarf candidates in
the \JHK\ color-color diagram is intriguing in light of our modeling
results.  (1)~Many of the objects exhibit $H\!-\!K$ excesses of several
tenths of a mag, substantially larger than the maximum excess predicted
by simple flat disk models (Table~\ref{table-passive}).  One possible
explanation is that these are flared disks seen close to edge-on; models
which include the effects of scattering show that such disks tend to
have redder colors than simple flat reprocessing disks
\citep{2002ApJ...567.1183W}.  If this is the case, the central sources
will be more heavily extincted, and since the target selection was based
on magnitudes, will tend to have higher masses.  Another possibility is
that accretion luminosity is a significant component of the disk
emission, \ie, $\dot M \gtrsim 10^{-9}$~\Msun~yr\perone\
(equation~(\ref{eqn:lumacclumrep})). In this case, the implication would
be much higher accretion rates for the Trapezium sample compared to the
Taurus and IC~348 objects in our sample, perhaps a reflection of
Trapezium's younger age and/or higher stellar density. (2)~Most of the
Trapezium IR excess sources lie below the CTTS locus of
\citet{1997AJ....114..288M}, namely they have bluer $(J\!-\!H)$ colors.
Simple disk models can explain some of the objects as being mid-M type
objects with disks, which will have bluer $J\!-\!H$ colors than the
early-M type objects used to define the CTTS locus (\eg,
Figure~\ref{plot-passive-2panel}).  However, these models cannot explain
sources with very blue colors ($J\!-\!H\lesssim0.6$ mag).  Muench \etal\
point out that half of these are associated with proplyds and hence are
genuine disk-bearing sources; perhaps their blue colors arise from
scattering.  The nature of the remaining blue sources remains
unresolved.

\citet{2002A&A...382L..22O} have studied the disk fraction for low mass
stars and a few brown dwarf candidates in the $\sigma$~Ori cluster.  For
a sample of 34 cluster members selected on the basis of optical
photometry and Li~absorption, they found at most 2 out of 34
($6\%\pm4\%$) objects had IR excesses, based on \JHK\ colors.  Adopting
a cluster age of $\sim$2--7~Myr, Oliveira \etal\ suggest that disks
around very low mass objects can dissipate within a few Myr.  These
conclusions are cast into doubt by our results.  We have shown that
\JHK\ colors are poor diagnostics of disks around VLM stars and brown
dwarfs.  Longer wavelength observations, at least in the $L$-band, are
needed to measure the true disk fraction and its implications for the
evolutionary time scale of disks around very low mass objects.


\subsection{Properties of Young Brown Dwarfs and their
   Circum(sub)stellar Disks \label{sec:disks}} 

Our results suggest that much of the observational paradigm developed
for disks around T~Tauri stars can be extended to disks around young
brown dwarfs.  Most of our sample show both IR excesses and strong
H$\alpha$ emission, like the higher mass classical T~Tauri stars.  Some
objects have little/no IR excesses and little H$\alpha$ emission,
analogous to weak-line T~Tauri stars. A few have IR excesses but only
weak H$\alpha$ emission; such a phenomenon is also seen among weak-line
T~Tauri stars \citep[e.g.][]{1989AJ.....97.1451S, 1990AJ.....99.1187S,
2002AJ....123.1613P}.  In addition, the widespread presence of H$\alpha$
emission suggests that most brown dwarfs are accreting, although the
inferred rates are much lower than classical T~Tauri stars.

We have also found evidence that young brown dwarfs, like their more
massive counterparts, the T~Tauri stars, possess circumstellar disks
with inner holes (\S~\ref{sec:holes}).  What is the origin of these
holes?  In the case of T~Tauri stars, inner holes are believed to be
created through the truncation of the disk by strong, closed stellar
magnetic fields; disk matter may reach the star by accretion along these
stellar magnetic field lines \citep[e.g.][]{1988ApJ...330..350B,
1991ApJ...370L..39K, 1994ApJ...426..669H, 1994ApJ...429..781S}.  A
similar situation may be relevant to young brown dwarfs.  Like their
more massive counterparts, young brown dwarfs also show signs of surface
magnetic activity.  For example, young brown dwarfs have detectable
X-ray emission that is characterized by plasma temperatures of
$\sim$1$-$2~keV and X-ray luminosities that are $\sim 10^{-4}$ to
$10^{-3}$ times the bolometric luminosity.  The X-ray emission also
displays moderate time variability and rapid flaring
\citep{2001ApJ...557..747I, 2001AJ....122..866P, 2002AJ....123.1613P}.
These similarities strongly suggest that young brown dwarfs have
significant magnetic fields.\footnote{This evidence for strong magnetic
activity among young brown dwarfs contrasts strongly with the evidence
for decreased magnetic activity among dwarfs at the bottom of the main
sequence \citep{2000AJ....120.1085G}.  This is likely due to the higher
temperatures of young brown dwarfs (i.e., spectral types of mid-M and
later), which are high enough to support a partially ionized atmosphere,
and therefore, surface magnetic fields, chromospheres, and coronae
\citep[e.g.][]{2002AJ....123.1613P}.}

The likely disk accretion rates of the brown dwarfs in our survey are
expected to be small (\S~\ref{sec:accretion}) and only modest magnetic
fields are needed to produce the inner holes with radii of 2$-$3~$R_*$.
For example, with typical parameters of $M_*=0.06~\Msun$,
$R_*=0.6~\Rsun$, and a disk accretion rate of
$10^{-10}$~\Msun~yr\perone, the required field strengths are $\sim
110-220$~G if the X-wind theory of \citet[][equation~2.6b with
$\alpha_X=1$]{1994ApJ...429..781S} is applied directly to this case (see
also \citealp{1995ApJ...447..813O} who recommend $\alpha_X=0.923$).
This estimate assumes that the magnetosphere is only partially filled,
with the accretion rate onto the star restricted to stellar loops that
correspond to $\sim$10\% of a filled magnetosphere
\citep{1998ApJ...509..802C, 2000ApJ...545L.141M}.  For comparison, the
formulation given by \citet{1991ApJ...370L..39K} when applied directly
to this case requires field strengths of $\sim 450-900$~G.  These field
strengths are quite modest, less than those measured for
pre-main-sequence stars \citep{1999A&A...341..768G, 1999ApJ...510L..41J,
1999ApJ...516..900J} and dMe stars \citep[e.g.][]{1996ApJ...459L..95J}.

It is interesting that the temperature at an inner disk radius of $2R_*$
in our disk models is approximately 1200~K, very similar to the
estimated temperature for the inner disk radii of T~Tauri stars
\citep[e.g.][]{1994ApJ...429..781S}.  At the densities of inner disks
and temperatures of $\gtrsim$1000~K, thermal ionization (of Na and K)
can maintain an ionization fraction that enables magnetic coupling
between the star (or brown dwarf) and the disk \citep{ume88,
2002MNRAS.329...18F}.  Therefore, the inner regions of brown dwarf disks
will be sufficiently ionized for magnetic fields to couple to and
truncate the disk at $\approx2\Rstar$.


\subsection{Implications for Brown Dwarf Formation}

For a star like the Sun, significant disk accretion is an imperative
because the angular momentum of the parent cloud core is typically too
large for the star to accrete more than a few percent of its final mass
through direct infall \citep[e.g.][]{1993prpl.conf....3S}.  If Sun-like
stars and brown dwarfs form under similar initial conditions (e.g., they
form from similar cloud cores), then the need for significant disk
accretion is much reduced since brown dwarfs may be able to accrete a
substantial fraction of their final mass through direct infall.  Our
results indicate that if brown dwarfs form through such means then disks
are produced.  In such a scenario, an important question is what
determines the final mass of the object, i.e., why a given core would
create a brown dwarf rather than a Sun-like star.

One physical effect which may address this concern is the premature
truncation of accretion due to strong dynamical interactions at early
times.  \citet{2001AJ....122..432R} have proposed that brown dwarfs
originate as stellar embryos in small multiple systems which are
prematurely ejected.  These embryos accrete from a common infalling
envelope, but dynamical interactions among them lead to the preferential
ejection of the lowest mass members.  Here, the final masses of the
ejected objects are established by the termination of accretion at the
time of ejection.  (Note that although their work focuses on brown
dwarfs, there is no clear distinction across the substellar boundary and
hence this scenario might also apply to VLM stars.)  The process is
purported to occur at very young ages ($\sim$10$^4$~yr), due to the
short crossing times of the multiple systems.  Preliminary studies of
young star-forming regions do not find any obvious kinematic difference
between the brown dwarfs and T~Tauri stars \citep{2001A&A...379L...9J},
as might be expected from preferential ejection of the young brown
dwarfs. However, \citet{2001astro.ph.10481R} suggest that the dynamical
signature of this process may be subtle and largely independent of mass.

An alternative class of formation scenarios invokes circumstellar disks
as the birthplace of substellar objects, perhaps through collisions of
star+disk systems or intrinsic instabilities in single star+disk systems
\citep[e.g.][]{1998Sci...281.2025L, 1998MNRAS.300.1205W,
1998MNRAS.300.1214W, 1998ApJ...503..923B}.  Specific predictions are
lacking; \eg, for simulations of star+disk collisions, the masses of the
resulting objects depend on the assumed star+disk properties and
especially the orbital parameters.  Nevertheless, dynamical
interactions, either during the formation episode or afterwards, are
needed to liberate the brown dwarfs into free-floating objects. In this
respect, similarities exist with the aforementioned
\citet{2001AJ....122..432R} scenario.  In a broader sense, all these
scenarios have randomness as a central deterministic element, as opposed
to, \eg, the initial conditions in pre-(sub)stellar molecular cores.

Intuition suggests that ejection will be inherently hostile to disks
around young substellar objects.  Studies of the encounter of a
star+disk system with another passing system find that such events are
highly disruptive to the disks, typically truncating the disk sizes to
half the periastron distance \citep{1993MNRAS.261..190C,
1996MNRAS.278..303H}. A similar outcome is seen by
\citet{1997MNRAS.285..540A} for the ejection of a star+disk system from
a small stellar cluster.  These effects are likely to be more severe for
disks around brown dwarfs given the shallower gravitational potential.
Furthermore, such encounters are likely to trigger rapid accretion
\citep{1992ApJ...401L..31B}.  Hence, \citet{2000AJ....120.3177R} and
\citet{2001AJ....122..432R} predict that young brown dwarfs are unlikely
to have signatures of disks, or else that such signatures will have
shorter lifetimes than for young stars.  This is inconsistent with our
results: we find that disks around young brown dwarfs are very common.
Moreover, the disks are contemporaneous with disks around the young
stars in the same star-forming regions. Assuming the brown dwarfs and
stars are roughly coeval, the lifetimes of brown dwarf disks are at
least as long as those of stellar disks.

One possible explanation for this discrepancy could be that most of the
(diskless) ejected objects have dispersed to large distances from the
young stars, which would occur for ejection velocities of a few~\kms.
Such objects would then be missing from existing optical/IR imaging
surveys.  The remaining brown dwarfs would be the ones which experienced
only very gentle ejections, and hence their disks would be preserved.
It is not clear if such a direct correspondence between ejection
velocity and disk survival can be maintained.  Moreover, in such a
scenario the mass function measured for IC~348 would be strongly
deficient in brown dwarfs compared to the field mass function.  This is
not supported by the observed IMF \citep{1999ApJ...525..466L,
2000ApJ...541..977N}.

Another possible explanation is that the post-ejection disks are
truncated but they accrete at such low rates that they still retain
enough disk material to produce an observable disk signature at ages of
a few~Myr. Detailed predictions for the effect of ejection on brown
dwarf disks are not yet available, but simple estimates suggest that the
surviving disks may be quite tenuous. We consider the case where the
velocities of the ejected brown dwarfs are not very different than the
stars, consistent with simulations of decaying few-body systems
\citep{1998A&A...339...95S}.  For a 50~\Mjup\ object, a typical velocity
of $\sim$2~\kms\ \citep{1979AJ.....84.1872J, 1986ApJ...309..275H}
corresponds to the escape velocity at a distance of 10~AU. Adopting this
estimate as the truncation radius, we can ask whether the disks will
have enough material to persist for a few~Myr. The minimum total disk
mass for the presence of optically thick \Lp-band emission is only
$\sim$10$^{-7}$~\Msun, which would quickly be accreted for canonical
accretion rates of 10$^{-10}$ to 10$^{-11}$~\Msun~yr\perone.  Instead
assume that the initial disk was 100~AU in size with a mass of 1\% of
the central object, similar to the case for T~Tauri stars
\citep{1995ApJ...439..288O}.  Assuming a surface density profile of
$r^{-3/2}$ \citep{1981PThPS..70...35H}, the truncated disk will have
(10/100)$^{1/2}$ of the initial mass, or a total mass of about
10$^{-4}$~\Msun\ for a 50~\Mjup\ object.  For the aforementioned
accretion rates, the lifetime of these truncated disks will be about
1--10~Myr, comparable to the ages of our targets. Hence, the disk mass
budget is roughly compatible with the ejection scenario.  Note that this
estimate is based on two assumptions: (1)~the current accretion rates
are roughly constant over the entire lifetime, and (2)~the disk surface
density profile is not perturbed significantly by the
ejection+truncation event.  Neither of these is likely to be realistic.

Alternatively, our finding that brown dwarf disks are as common as disks
around T~Tauri stars is well-accommodated in perhaps the simplest
picture of brown dwarf formation, namely the collapse of isolated, very
low mass cores --- a direct analog to the conventional picture of
isolated low-mass star formation.  In this case, the low disk accretion
rates found for young brown dwarfs may have played a more central role
in their formation.  That is, brown dwarfs that coexist with stars in
young clusters like IC~348 may have formed like their more massive
counterparts but with much lower accretion rates in order for all the
objects to have accumulated their final masses on roughly the same
timescale.  The idea that stars of different masses result from
different accretion rates is not new.  For example, something similar is
needed in order to explain the coexistence of high mass stars (e.g.,
$>$20~\Msun) and low mass stars (e.g., $< 1 M_\odot$) in young clusters
\citep[e.g.][]{1987ARA&A..25...23S, 1993ApJ...402..635M}.  Since the
main sequence lifetime of high mass stars is only a few Myr, they must
accrete their final mass on a timescale much shorter than this in order
to coexist with much lower mass stars.  The mass accretion rates
required ($\sim 10^{-4}$~\Msun~yr\perone) are much larger than those
typically ascribed to the young progenitors of solar-mass stars.  Our
results suggest that a similar scenario of mass-dependent accretion
rates may apply down to substellar masses.

To summarize, the widespread presence of disks around young brown dwarfs
at ages of a few~Myr provides prima facie evidence of a similar origin
for low-mass stars and brown dwarfs.  The high disk frequency is more
difficult to reconcile with scenarios involving disk-disk collisions
and/or premature ejection.  Important future observational goals will be
to determine the sizes and masses of brown dwarf disks.  These cannot be
measured from existing thermal-IR ($\sim$3$-$10~\micron) data, since
emission at these wavelengths originates from relatively small radii
($\lesssim$1~AU) and is optically thick even for minute disk masses.
Longer wavelength data, namely far-IR (from \SIRTF) and sub-mm (\eg,
from JCMT/SCUBA and SMA) measurements are needed.


\section{Summary and Concluding Remarks}

We have completed the first systematic survey for circumstellar disks
around young brown dwarfs and VLM stars.  We have obtained \Lp-band
imaging for objects spectroscopically classified as having very low
temperatures and combined these observations with existing \JHKs\ data.
Our sample comprises most of the published objects with spectral types
from M6 to M9.5 in IC~348 and Taurus, and should be largely free of
selection biases for our purposes.  None appear to be binaries at a
resolution of 0\farcs4 resolution (55--120~AU).  Based on current
models, our targets have masses of $\sim$15 to 100~\Mjup\ and ages of
$\lesssim$5~Myr.
Our survey is sensitive enough to detect bare substellar photospheres
and hence is the first unbiased survey for disks around such low mass
objects. 

Using the published spectral types and extinctions, we determine the
intrinsic photospheric \KmLp\ (2.0$-$4.1~\micron) colors of our targets
and hence if they have IR excesses.  We find that excesses are very
common, occurring in nearly 80\% of the sample.  Such excesses are one
of the classic signatures of circumstellar disks around T~Tauri stars,
and there is ample observational evidence and theoretical expectation
that such disks exists around young stars.  Likewise, disks are the
natural and most plausible explanation for the IR excesses we observe
around young brown dwarfs and VLM stars.  The \Lp-band emission arises
from the inner regions of the disks, within $\lesssim$0.1~AU, and is
sensitive to disks with masses of $\gtrsim10^{-7}$~\Msun.

The observed \KmLpo\ excesses are well-correlated with H$\alpha$ line
emission, as in the case for T~Tauri stars.  This supports the idea that
accretion disks are the common origin for both forms of emission.  The
IR excesses do not appear to correlate with any other property of the
central sources.  The excess frequency and amplitude are independent of
the mass.  However, the coolest objects in the sample, spectral type M9,
have negligible excesses, indicating that their inner disk regions are
empty or perhaps the objects are diskless.  Given the very low estimated
masses of the M9 objects ($\lesssim15-20$~\Mjup), this is a tantalizing
finding which should be pursued with a larger sample.

The T~Tauri stars of IC~348 and Taurus exhibit comparably high disk
fractions as the brown dwarfs, and therefore we find that stellar and
substellar disks are contemporaneous.  Assuming the populations in these
regions are roughly coeval, brown dwarf disks are at least as long-lived
as disks around solar-mass stars.  The amplitude of the IR excesses do
not show any trend with age, using the absolute $K$-band magnitude as a
model-independent relative age scale.  Most of the stellar population
has an estimated age of 1--3~Myr, and assuming coevality, we infer that
the inner regions of brown dwarf disks do not evolve significantly for
the first $\sim$3~Myr.

Given the low luminosities and weak gravitational potentials of
substellar objects, one expects a priori that circumstellar disks around
young brown dwarfs will be cooler and less luminous than disks around
the higher mass T~Tauri stars.  This is confirmed by our measurements:
the observed \KmLpo\ excesses are substantially smaller than those for
T~Tauri stars.  As a consequence, a conventional analysis of the disk
fraction using only IR colors would have missed most of the sources with
excesses in our sample.  While such analyses work well for T~Tauri
stars, in the case of young brown dwarfs the much reduced contrast
between the disks and the photospheres is a serious impediment.

We use standard models of flat, optically thick circumstellar disks to
examine the physical properties of the brown dwarf disks.  The IR
excesses can be explained simply by reprocessing disks --- there is no
need for a significant contribution from accretion luminosity. The
inferred accretion rates are $\lesssim10^{-9}$~\Msun~yr\perone, or at
least an order of magnitude lower than typical for classical T~Tauri
stars.  Nevertheless, the presence of H$\alpha$ emission strongly
suggests that at least some accretion is ongoing.  The mere existence
of accreting brown dwarfs at ages of a few Myr argues for mass-dependent
accretion rates, since brown dwarfs with typical T~Tauri star accretion
rates would not remain substellar.

The maximum observed excesses are less than those expected from a disk
which extends to the stellar surface --- this suggests the disks have
inner holes.  This inference is supported by analyzing the observed
distribution of IR excesses with Monte Carlo simulations of disks viewed
at random inclinations: models with hole sizes of $\approx2\Rstar$ are a
much better match to the observations than models without holes or those
with much larger holes.  The presence of an inner hole is consistent
with a picture where the hole arises from the interaction of the inner
disk with the magnetic field of the central object.  Our rough estimates
based on magnetospheric accretion models find that the brown dwarf
magnetic fields are on the order of several hundred gauss, quite modest
compared to measurements for T~Tauri stars.  The inner hole sizes for
both T~Tauri stars and young brown dwarfs are indicative of disk
temperatures hot enough for significant thermal ionization, which is
needed to permit coupling of the magnetic field to the disk.  Thus, our
inferred hole sizes suggest that the magnetic accretion paradigm
developed for T~Tauri stars may extend to substellar masses.  This also
illustrates one way in which brown dwarf disks may be useful
laboratories for testing our understanding of the physical processes of
circumstellar disks.

We have quantitatively examined the possible systematic errors and find
that they are not expected to be significant.  These include the
uncertainties in spectral typing and extinction, disk emission
contamination, surface gravity effects, sample selection bias, and the
temperature scale for pre-main sequence objects.  In addition, most of
the plausible systematic effects, even if significant, would act to
increase the disk fraction, which we already find is very high.

Our basic observational findings are that (1)~most young brown dwarfs
have disks, and (2)~these disks are contemporaneous with disks around
T~Tauri stars in the same star-forming regions.  The latter also
demonstrates that brown dwarf disks are at least as long-lived as disks
around stars, assuming that the stars and brown dwarfs are roughly
coeval.  These observations are naturally accommodated in a picture
where brown dwarfs are born in a similar manner as stars --- our results
offer compelling evidence for a common origin for objects from the
stellar regime down to the substellar and planetary-mass regime.

Alternative formation scenarios, such as disk-disk collisions and
premature ejection, involve dynamical interactions in creating brown
dwarfs.  While specific predictions are lacking due to the stochastic
nature of these scenarios, brown dwarfs formed by collision and/or
ejection are generally expected to have smaller and less massive disks,
and consequently shorter disk lifetimes, compared to brown dwarfs formed
in isolation.  This expectation conflicts with our finding that brown
dwarf disks are at least as long-lived as disks around young stars.

We have found that brown dwarf disks are common.  In order to better
understand the nature of these disks, longer wavelength (far-IR, sub-mm,
and mm) measurements are needed to determine the disk masses and sizes.
More measurements of the accretion rates are sorely needed.  The
lifetimes of disks around brown dwarfs and the disk frequency at even
lower masses ($\lesssim$15~\Mjup) are also outstanding questions.  These
studies will be important for determining the origin of brown dwarfs and
their disks, and also for placing these objects in context with our
physical understanding of the star formation process.  Finally, the very
high disk fraction of young brown dwarfs raises the possibility of
forming planets around brown dwarfs.  Such planetary systems would
represent a fascinating alternative to the numerous planetary systems
found around solar-type stars.


\acknowledgments

It is a pleasure to thank Charlie Lada, James Muzerolle, John Tonry, and
Jonathan Williams for useful discussions.  We are grateful to support
from the staff of UKIRT observatory, especially Thor Wold and Paul
Hirst, which  made these observations possible.  We thank Sandy Leggett
for updated information on the standard stars and for making filter
curves and her published photometry readily available.  We thank Kevin
Luhman for providing unpublished H$\alpha$ measurements.  UKIRT is
operated by the Joint Astronomy Centre on behalf of the U.K.\ Particle
Physics and Astronomy Research Council.  This research has made use of
data products from the Two Micron All Sky Survey, which is a joint
project of the University of Massachusetts and the Infrared Processing
and Analysis Center/California Institute of Technology, funded by NASA
and NSF.  This research has also made use of NASA's Astrophysics Data
System Abstract Service.  M.~Liu is grateful for research support from
the Beatrice Watson Parrent Fellowship at the University of Hawai`i.



\appendix
\section{Calculating the Disk Fraction in the Presence of Observational Errors}

\newcommand{\Ndisk}{\mbox{$N_{disk}$}}
\newcommand{\fdisk}{\mbox{$f_{disk}$}}
\newcommand{\sigfdisk}{\mbox{$\sigma(f_{disk})$}}
\newcommand{\calG}{\mbox{$\cal G$}}
\newcommand{\calP}{\mbox{$\cal P$}}

In the literature, the usual means of computing the disk fraction
assumes only Poisson counting statistics contribute to the uncertainty
in the result. Consider a sample of $N$ objects, each with a color
excess $\Delta_i$, which can be derived from a single color (as in the case
of our work here) or many colors (\eg, a color-color diagram).
Typically, one then defines the number of sources with disks as
\begin{equation}
\Ndisk = \sum_i{(\Delta_i>\Delta_0)}
\end{equation}
where the notation represents the number of measurements greater than a
specified constant $\Delta_0$.  Many studies choose $\Delta_0=0$, \ie,
any color excess is taken as the sign of a disk
\citep[e.g.][]{1995AJ....109.1682L, 2001AJ....121.2065H}. The disk
fraction and its error are then
\begin{equation}
\fdisk = {\Ndisk \over N}
\end{equation}
\begin{equation}
\sigfdisk = {\sqrt{\Ndisk} \over N}\ ,
\label{eqn:sigfdisk}
\end{equation}
respectively. Clearly, this approach neglects the effects of measurement
errors: a sample with very large errors in the $\Delta_i$'s should lead
to a larger \sigfdisk\ than a sample with small errors.  Only in the
case of zero measurement errors will \sigfdisk\ be given exactly by
equation~(\ref{eqn:sigfdisk}).  Hence, the published disk fraction errors
are somewhat underestimated.  Some studies attempt to circumvent the
influence of observational errors by setting $\Delta_0>0$, typically
choosing a value comparable to the measurement errors
\citep[e.g.][]{1989AJ.....97.1451S, 1998AJ....116.1816H}. However, this
approach inevitably excludes some objects with genuine disk excesses,
and hence the disk fractions are somewhat
underestimated.

We use a simple Monte Carlo approach to account for both Poisson
counting errors and measurement errors in determining the disk
fraction.\footnote{\citet{2001AJ....121.2673K} use a Monte Carlo
approach to account for measurement errors and uncertainty in the
extinction law when computing the disk fraction. However, they neglect
Poisson errors in their computations.}  Our observed sample is composed
of $N$ objects, each with a color index $\Delta_i$ and measurement error
$\epsilon_i$.  For each Monte Carlo realization, we generate a set of
$N$ objects having color indices
\begin{equation}
\Delta_i^\prime = \Delta_i + (\calG_i \times \epsilon_i)
\end{equation}
where $\calG_i$ is drawn from a normal distribution of zero mean and
unit variance. We then determine the number of sources with
$\Delta_i^\prime>0$, denoted as $\overline{N}_{disk}^{\,j}$.  (The
subscript~$i$ denotes different objects, and the superscript~$j$ denotes
different Monte Carlo realizations.)  For each realization, we assume
the resulting number of sources with disks is subject to counting
statistics: namely, the number of disk sources is described by a Poisson
distribution of mean $\overline{N}_{disk}^{\,j}$, which we denote as
$\calP(N_{disk}^j; \overline{N}_{disk}^{\,j})$.  After a large number of
realizations (we use 100,000), we sum all the individual Poisson
distributions to form a probability distribution function for the number
of disk sources:
\begin{equation}
{\rm PDF}(\Ndisk) = \sum_j \calP(N_{disk}^j; \overline{N}_{disk}^{\,j})\ .
\end{equation}
We then compute the disk fraction using the value of \Ndisk\ at the peak
of the PDF and use the standard deviation of the PDF (well approximated
by fitting a gaussian) as the error in \Ndisk.

The net result of using this Monte Carlo approach is that our tabulated
disk fractions are $\approx$5\% smaller than simply using the standard
approach (\ie, equation~(A1) with $\Delta_0=0$) and the errors in the
disk fraction are $\approx$5\% larger.  The effects are small because of
the relatively high S/N of the \KmLpo\ excess measurements (median error
of 0.09~mag).  Note that these percentages should not be used as a
general representation for all disk fraction calculations.  The result
of ignoring the measurement errors in $\Delta_i$ will depend on a
particular sample's properties, namely the size of the sample, the
distribution of excesses, and the measurement errors.

\setcounter{section}{0}







\clearpage


\clearpage

\clearpage

\clearpage

\begin{deluxetable}{lcccc}
\tablecaption{Infrared Photometry \label{mags-final}}
\tablewidth{0pt}
\tablehead{
  \colhead{Object} &
  \colhead{Spectral Type} &
  \colhead{$A_K$ (mag)} &
  \colhead{\Lp\ (mag)} &
  \colhead{$E(\KmLp)_0$ (mag)}
}
\startdata

\cutinhead{IC 348: core sample}
  NTC 075-07  &  M5.7 $\pm$ 0.6 &   0.00 $\pm$ 0.11 &	 13.03 $\pm$ 0.12 &  \phs0.06 $\pm$ 0.15 \\
  NTC 071-01  &  M5.8 $\pm$ 0.7 &   0.04 $\pm$ 0.08 &	 13.42 $\pm$ 0.05 &  \phs0.33 $\pm$ 0.10 \\
  NTC 025-04  &  M5.9 $\pm$ 0.6 &   0.00 $\pm$ 0.09 &	 11.86 $\pm$ 0.06 &  \phs0.03 $\pm$ 0.10 \\
  NTC 053-03  &  M5.9 $\pm$ 0.6 &   0.51 $\pm$ 0.06 &	 12.36 $\pm$ 0.03 &   $-$0.20 $\pm$ 0.07 \\
  NTC 053-01  &  M6.1 $\pm$ 0.6 &   0.18 $\pm$ 0.06 &	 10.75 $\pm$ 0.04 &  \phs0.45 $\pm$ 0.08 \\
  NTC 044-04  &  M6.2 $\pm$ 0.6 &   0.02 $\pm$ 0.06 &	 11.82 $\pm$ 0.04 &  \phs0.15 $\pm$ 0.08 \\
  NTC 055-02  &  M6.3 $\pm$ 0.6 &   0.24 $\pm$ 0.06 &	 12.19 $\pm$ 0.04 &  \phs0.29 $\pm$ 0.08 \\
  NTC 014-05  &  M6.3 $\pm$ 0.6 &   0.04 $\pm$ 0.07 &	 12.96 $\pm$ 0.05 &  \phs0.26 $\pm$ 0.09 \\
  NTC 011-02  &  M6.4 $\pm$ 0.6 &   0.00 $\pm$ 0.18 &	 12.29 $\pm$ 0.05 &   $-$0.04 $\pm$ 0.13 \\
  NTC 075-06  &  M6.4 $\pm$ 0.7 &   0.06 $\pm$ 0.06 &	 12.45 $\pm$ 0.07 &  \phs0.24 $\pm$ 0.10 \\
  NTC 051-01  &  M6.5 $\pm$ 0.6 &   0.17 $\pm$ 0.06 &	 11.66 $\pm$ 0.04 &  \phs0.03 $\pm$ 0.08 \\
  NTC 062-01  &  M6.8 $\pm$ 0.7 &   0.19 $\pm$ 0.08 &	 13.15 $\pm$ 0.08 &  \phs0.13 $\pm$ 0.11 \\
  NTC 014-04  &  M6.8 $\pm$ 0.6 &   0.22 $\pm$ 0.06 &	 11.06 $\pm$ 0.04 &  \phs0.16 $\pm$ 0.08 \\
  NTC 022-05  &  M6.8 $\pm$ 0.6 &   0.53 $\pm$ 0.07 &	 12.42 $\pm$ 0.04 &  \phs0.24 $\pm$ 0.08 \\
  NTC 045-02  &  M7.1 $\pm$ 0.6 &   0.00 $\pm$ 0.10 &	 12.59 $\pm$ 0.04 &   $-$0.31 $\pm$ 0.09\tablenotemark{a} \\
  NTC 062-03  &  M7.2 $\pm$ 0.6 &   0.00 $\pm$ 0.15 &	 12.75 $\pm$ 0.06 &  \phs0.02 $\pm$ 0.12 \\
  NTC 011-01  &  M7.2 $\pm$ 0.6 &   0.02 $\pm$ 0.06 &	 12.69 $\pm$ 0.04 &   $-$0.04 $\pm$ 0.08 \\
  NTC 013-05  &  M7.5 $\pm$ 0.6 &   0.05 $\pm$ 0.07 &	 12.48 $\pm$ 0.04 &  \phs0.01 $\pm$ 0.08 \\
  NTC 042-03  &  M7.6 $\pm$ 0.6 &   0.10 $\pm$ 0.07 &	 12.22 $\pm$ 0.06 &   \phs0.00 $\pm$ 0.09 \\
  NTC 013-06  &  M7.7 $\pm$ 0.7 &   0.26 $\pm$ 0.07 &	 12.67 $\pm$ 0.04 &  \phs0.30 $\pm$ 0.09 \\
  NTC 043-06  &  M8.2 $\pm$ 0.6 &   0.06 $\pm$ 0.07 &	 11.80 $\pm$ 0.04 &  \phs0.31 $\pm$ 0.08 \\
  NTC 105-01  &  M8.2 $\pm$ 0.7 &   0.01 $\pm$ 0.07 &	 12.58 $\pm$ 0.14 &  \phs0.16 $\pm$ 0.16 \\
  NTC 015-06  &  M9.1 $\pm$ 0.7 &   0.00 $\pm$ 0.09 &	 12.76 $\pm$ 0.13 &  \phs0.02 $\pm$ 0.16 \\
  NTC 012-02  &  M9.4 $\pm$ 0.9 &   0.31 $\pm$ 0.11 &	 14.08 $\pm$ 0.09 &   $-$0.03 $\pm$ 0.14 \\
     
\cutinhead{IC 348: outer sample}     
     L99 205  &  M6.0 $\pm$ 0.2 &   0.13 $\pm$ 0.04 &	 11.43 $\pm$ 0.04 &  \phs0.38 $\pm$ 0.06 \\
     L99 312  &  M6.0 $\pm$ 0.2 &   0.20 $\pm$ 0.04 &	 12.60 $\pm$ 0.08 &   $-$0.14 $\pm$ 0.09 \\
     L99 325  &  M6.0 $\pm$ 0.2 &   0.19 $\pm$ 0.04 &	 12.27 $\pm$ 0.05 &  \phs0.39 $\pm$ 0.07 \\
     L99 367  &  M6.5 $\pm$ 0.5 &   0.05 $\pm$ 0.04 &	 12.86 $\pm$ 0.12 &  \phs0.16 $\pm$ 0.14 \\
     L99 382  &  M6.5 $\pm$ 0.5 &   0.30 $\pm$ 0.04 &	 12.69 $\pm$ 0.07 &  \phs0.37 $\pm$ 0.10 \\
     L99 407  &  M7.0 $\pm$ 0.5 &   0.40 $\pm$ 0.20 &	 13.77 $\pm$ 0.04 &  \phs0.11 $\pm$ 0.17 \\
     L99 363  &  M8.0 $\pm$ 0.2 &   0.00 $\pm$ 0.04 &	 13.05 $\pm$ 0.05 &  \phs0.06 $\pm$ 0.07 \\
     L99 405  &  M8.0 $\pm$ 0.2 &   0.00 $\pm$ 0.04 &	 13.20 $\pm$ 0.07 &  \phs0.15 $\pm$ 0.09 \\
     
\cutinhead{Taurus}     
V410 Anon 13  &    M6 $\pm$ 1.0 &   0.42 $\pm$ 0.06 &	 10.00 $\pm$ 0.04 &  \phs0.25 $\pm$ 0.11 \\
 V410 Xray 3  &    M6 $\pm$ 0.2 &   0.09 $\pm$ 0.06 & \phn9.79 $\pm$ 0.04 &  \phs0.10 $\pm$ 0.07 \\
       MHO 5  & M6.25 $\pm$ 0.2 &   0.01 $\pm$ 0.06 & \phn9.28 $\pm$ 0.04 &  \phs0.30 $\pm$ 0.07 \\
  CFHT-Tau 1  &    M7 $\pm$ 0.5 &   0.30 $\pm$ 0.10 &	 11.14 $\pm$ 0.08 &  \phs0.02 $\pm$ 0.11 \\
  CFHT-Tau 4  &    M7 $\pm$ 0.5 &   0.30 $\pm$ 0.10 & \phn9.15 $\pm$ 0.03 &  \phs0.47 $\pm$ 0.09 \\
  CFHT-Tau 2  &    M8 $\pm$ 0.5 &   0.00 $\pm$ 0.10 &	 11.41 $\pm$ 0.06 &  \phs0.16 $\pm$ 0.10 \\
  CFHT-Tau 3  &    M9 $\pm$ 0.5 &   0.00 $\pm$ 0.10 &	 11.63 $\pm$ 0.03 &  \phs0.00 $\pm$ 0.09 \\

\enddata

\tablenotetext{a}{This object has $\KmLp=0.23\pm0.04$, much bluer than
the other objects in the sample.  Based on its \JHKsLp\ colors, it
appears to be a background late-type giant star.  It is excluded from
the analysis.}

\end{deluxetable}

\clearpage

\begin{deluxetable}{cccc}
\tablecaption{Fraction of Sources with IR Excesses \label{table-excess}}
\tablewidth{0pt} 
\tablehead{ 
  \colhead{Spectral} & 
  \colhead{IC 348} &
  \colhead{IC 348} & 
  \colhead{Total}\\ 
  \colhead{Type} & 
  \colhead{core} &
  \colhead{core + outer} & 
  \colhead{(IC 348 + Taurus)}
} 
\startdata 
 5.7 -- 6.4 &  8/10  & 10/13 & 13/16 \\ 
 6.5 -- 7.4 &  5/6   & 8/9   & 10/11 \\ 
 7.5 -- 8.4 &  4/5   & 6/7   & 7/8 \\ 
 8.5 -- 9.4 &  1/2   & 1/2   & 1/3 \\
 all types  &  18/23 & 25/31 & 31/38 \\
            &  ($73\%\pm19\%$) & ($75\%\pm17\%$) & ($77\%\pm15\%$) \\
\enddata

%

\tablecomments{The percentages of sources with IR excesses and their
errors were evaluated using a Monte Carlo technique which accounts for
both measurement errors and Poisson counting statistics. See Appendix~A.}

\end{deluxetable}


\begin{deluxetable}{cccccc}
\tablecaption{Flat Blackbody Reprocessing Disk: Maximum Excess (in mags)
              \label{table-passive}} 
\tablewidth{0pt}
\tablehead{
  \colhead{Spectral Type} &
  \colhead{$\Delta J$} &
  \colhead{$\Delta H$} &
  \colhead{$\Delta \Ks$} &
  \colhead{$\Delta \Lp$} &
  \colhead{$\Delta \Mp$} 
}
\startdata

   M0  &  0.27  &  0.38  &  0.63  &  1.38  &  1.78  \\
   M1  &  0.23  &  0.35  &  0.60  &  1.32  &  1.72  \\
   M2  &  0.20  &  0.33  &  0.56  &  1.26  &  1.66  \\
   M3  &  0.18  &  0.31  &  0.52  &  1.20  &  1.59  \\
   M4  &  0.16  &  0.28  &  0.49  &  1.13  &  1.52  \\
   M5  &  0.14  &  0.26  &  0.45  &  1.05  &  1.44  \\
   M6  &  0.12  &  0.24  &  0.42  &  0.98  &  1.36  \\
   M7  &  0.10  &  0.22  &  0.38  &  0.89  &  1.27  \\
   M8  &  0.09  &  0.19  &  0.34  &  0.81  &  1.17  \\
   M9  &  0.08  &  0.17  &  0.31  &  0.73  &  1.08  \\

\enddata

\tablecomments{Transmission profiles for the \JHKs\ bands are from
2MASS, while those for the \Lp\ and \Mp\ bands are from the MKO system.}

\end{deluxetable}

\end{document}